\documentclass{article}

\usepackage[eandd,preprint]{neurips_2026}

\usepackage[utf8]{inputenc} %
\usepackage[T1]{fontenc}    %
\usepackage{url}            %
\usepackage{amsfonts}       %
\usepackage{nicefrac}       %
\usepackage{microtype}      %

\usepackage{xspace}
\usepackage{xcolor}
\usepackage{graphicx}
\usepackage[normalem]{ulem} %
\usepackage{amsmath}
\usepackage{enumitem}

\usepackage{hyperref}
\hypersetup{linkcolor=black,citecolor=black,anchorcolor=black,filecolor=black,menucolor=black,runcolor=black,urlcolor=black,hidelinks}
\usepackage{breakurl}

\usepackage{booktabs}
\usepackage{multirow}
\usepackage{makecell}
\usepackage{ragged2e}

\usepackage{caption}
\usepackage{subcaption}
\usepackage{wrapfig} %

\usepackage{listings}

\usepackage{tikz}
\usetikzlibrary{calc}
\usetikzlibrary{shapes.geometric}
\usetikzlibrary{decorations.pathreplacing}
\usetikzlibrary{positioning}

\usepackage{datetime} %
\usepackage{tcolorbox} %

\makeatletter
\newcommand{\DefMacro}{\@ifstar\@DefMacroAllowRedefine\@DefMacro}
\newcommand{\@DefMacro}[2]{\expandafter\newcommand\csname rmk-#1\endcsname{#2}}
\newcommand{\@DefMacroAllowRedefine}[2]{\expandafter\providecommand\csname rmk-#1\endcsname{} \expandafter\renewcommand\csname rmk-#1\endcsname{#2}}
\makeatother
\newcommand{\UseMacro}[1]{\csname rmk-#1\endcsname}

\newcommand{\XSpace}[1]{}
\newcommand{\XComment}[1]{}

\newcommand{\MyPara}[1]{\noindent\textbf{#1}.}

\newcommand{\Code}[1]{{\ifmmode{\mathtt{#1}}\else$\mathtt{#1}$\fi}}
\newcommand{\CodeIn}[1]{{\ifmmode{\mathtt{#1}}\else$\mathtt{#1}$\fi}}

\newcolumntype{R}[1]{>{\RaggedLeft\arraybackslash}p{#1}}
\newcolumntype{L}[1]{>{\RaggedRight\arraybackslash}p{#1}}

\definecolor{gray}{RGB}{211,211,211}
\newcommand{\jbasicstyle}{\small\sffamily} %

\newcommand{\jnumberstyle}{\scriptsize}

\lstdefinelanguage{pseudo}
{
morekeywords={},
keywordstyle=\bfseries,
lineskip=-0.1em,
numbers=left, %
numberstyle=\jnumberstyle,
numbersep=4pt,
basicstyle=\jbasicstyle,
breaklines=true,
breakautoindent=true,
tabsize=2,
columns=fullflexible,
morecomment=*[l][\textsl]{//},
mathescape=true,
xleftmargin=10pt,
}

\lstdefinelanguage{todo-comment}
{
morekeywords={},
keywordstyle=\bfseries,
lineskip=-0.1em,
numbers=none,
basicstyle=\jbasicstyle,
breaklines=true,
breakautoindent=true,
tabsize=2,
columns=fullflexible,
morecomment=*[l][\textsl]{//},
mathescape=true,
xleftmargin=-10pt,
}

\lstdefinelanguage{java-pretty}
{
language=java,
numbers=left,
basicstyle=\scriptsize\ttfamily,
numberstyle=\scriptsize,
breaklines=true,
columns=fullflexible,
xleftmargin=16pt,
showstringspaces=false,
}

\newcommand{\Bench}{\textsc{TestEvo-Bench}\xspace}
\newcommand{\Title}{\Bench: An Executable and Live Benchmark for Test and Code Co-Evolution}
\newcommand{\URL}{https://www.testevo-bench.com} %

\newcommand{\SWEAgent}{SWE-Agent\xspace}
\newcommand{\ClaudeCode}{Claude Code\xspace}
\newcommand{\GeminiCLI}{Gemini CLI\xspace}
\newcommand{\OpenAICodexCLI}{OpenAI Codex CLI\xspace}
\newcommand{\GPTCodex}{GPT 5 Codex\xspace}
\newcommand{\TestUpdater}{TestUpdater\xspace}
\newcommand{\AgoneTest}{AgoneTest\xspace}
\newcommand{\ClaudeOpus}{Claude Opus 4.7\xspace}

\newcommand{\GeminiPro}{Gemini 3.1 Pro\xspace}
\newcommand{\Cursor}{Cursor\xspace}
\newcommand{\CursorComposer}{Composer 2.0\xspace}

\newcommand{\NumCandidateRecords}{59{,}950\xspace}
\newcommand{\NumCandidateProjects}{152\xspace}
\newcommand{\NumClassifiedRecords}{13{,}868\xspace}
\newcommand{\NumGenerationMethodsCurrent}{1{,}961\xspace}
\newcommand{\NumUpdateMethodsCurrent}{1{,}138\xspace}

\newcommand{\OldOnOld}{T\textsuperscript{old} \textrightarrow C\textsuperscript{old}}
\newcommand{\NewOnNew}{T\textsuperscript{new} \textrightarrow C\textsuperscript{new}}
\newcommand{\OldOnNew}{T\textsuperscript{old} \textrightarrow C\textsuperscript{new}}
\newcommand{\NewOnOld}{T\textsuperscript{new} \textrightarrow C\textsuperscript{old}}

\newcommand{\PassPct}{Pass\%\xspace}
\newcommand{\RedundantPct}{Redundant\%\xspace}
\newcommand{\ExecFailPct}{ExecFail\%\xspace}
\newcommand{\CmplFailPct}{CmplFail\%\xspace}
\newcommand{\HrnsFailPct}{HrnsFail\%\xspace}
\newcommand{\SuccessPct}{Success\%\xspace}
\newcommand{\CovMetric}{Cov\xspace}
\newcommand{\CovOnPassMetric}{CovOnPass\xspace}
\newcommand{\CovOnPassHumanMetric}{CovOnPassHuman\xspace}
\newcommand{\DeltaCovOnPassMetric}{$\Delta$CovOnPass\xspace}
\newcommand{\MutMetric}{Mut\xspace}
\newcommand{\MutOnPassMetric}{MutOnPass\xspace}
\newcommand{\MutOnPassHumanMetric}{MutOnPassHuman\xspace}
\newcommand{\DeltaMutOnPassMetric}{$\Delta$MutOnPass\xspace}

\DefMacro{TH-Harness}{Harness\xspace}
\DefMacro{TH-Model}{Model\xspace}
\DefMacro{TH-PassPct}{\PassPct}
\DefMacro{TH-RedundantPct}{\RedundantPct}
\DefMacro{TH-ExecFailPct}{\ExecFailPct}
\DefMacro{TH-CmplFailPct}{\CmplFailPct}
\DefMacro{TH-HrnsFailPct}{\HrnsFailPct}
\DefMacro{TH-SuccessPct}{\SuccessPct}
\DefMacro{TH-Cov}{\CovMetric}
\DefMacro{TH-CovOnPass}{\CovOnPassMetric}
\DefMacro{TH-CovOnPassHuman}{\CovOnPassHumanMetric}
\DefMacro{TH-DeltaCovOnPass}{\DeltaCovOnPassMetric}
\DefMacro{TH-Mut}{\MutMetric}
\DefMacro{TH-MutOnPass}{\MutOnPassMetric}
\DefMacro{TH-MutOnPassHuman}{\MutOnPassHumanMetric}
\DefMacro{TH-DeltaMutOnPass}{\DeltaMutOnPassMetric}

\DefMacro{RQ2-TUClaude-CompRate}{74.2\%}
\DefMacro{RQ2-TUGPT-CompRate}{62.5\%}
\DefMacro{RQ2-SAClaude-CompRate}{43.8\%}
\DefMacro{RQ2-SAGPT-CompRate}{48.1\%}
\DefMacro{RQ2-TUClaude-PassRate}{69.9\%}
\DefMacro{RQ2-TUGPT-PassRate}{56.2\%}
\DefMacro{RQ2-SAClaude-PassRate}{40.2\%}
\DefMacro{RQ2-SAGPT-PassRate}{43.5\%}
\DefMacro{RQ2-FrameworkGapPP}{21}
\DefMacro{RQ2-ModelGapTUPP}{14}
\DefMacro{RQ2-ModelGapSAPP}{3}
\DefMacro{RQ2-TUClaude-AvgCov}{80.0\%}
\DefMacro{RQ2-SAGPT-AvgCov}{71.2\%}
\DefMacro{RQ3-TotalCompileErrors}{3{,}175}
\DefMacro{RQ3-TotalTestErrors}{404}
\DefMacro{RQ3-CFS-Pct}{42.9\%}
\DefMacro{RQ3-CFS-Total}{1{,}363}
\DefMacro{RQ3-CFS-TU}{302}
\DefMacro{RQ3-CFS-SA}{1{,}061}
\DefMacro{RQ3-CFS-SA-vs-TU-Ratio}{3.5}
\DefMacro{RQ3-EmptyGenPct}{9.9\%}
\DefMacro{RQ3-EmptyGenTotal}{850}
\DefMacro{RQ3-BuildEnvPct}{10.9\%}
\DefMacro{RQ3-BuildEnvTotal}{346}
\DefMacro{RQ3-TypeAPIPct}{7.9\%}
\DefMacro{RQ3-AssertionErrPct}{24.3\%}
\DefMacro{RQ3-AssertionErrTotal}{98}
\DefMacro{RQ3-InitErrPct}{2.5\%}
\DefMacro{RQ3-InitErrTotal}{10}
\DefMacro{RQ3-ClassNotFoundPct}{8.2\%}
\DefMacro{RQ3-ClassNotFoundTotal}{33}
\DefMacro{RQ3-RuntimeExPct}{16.8\%}
\DefMacro{RQ3-TU-GPT-AssertionErr}{64}
\DefMacro{RQ3-SA-AssertionErr}{5}
\DefMacro{RQ3-NPE-Total}{24}
\DefMacro{RQ3-NPE-TU}{13}
\DefMacro{RQ3-NPE-SA}{11}
\DefMacro{RQ3-UnsolvableTotal}{306}
\DefMacro{RQ3-UnsolvablePct}{14.2\%}
\DefMacro{RQ3-Top3UnsolvableTotal}{110}
\DefMacro{RQ3-Top3UnsolvablePct}{35.9\%}
\DefMacro{RQ3-CompileBlockedPct}{62.1\%}
\DefMacro{RQ3-CompileBlocked}{190}
\DefMacro{RQ3-LogicBlockedPct}{37.9\%}
\DefMacro{RQ3-LogicBlocked}{116}

\DefMacro{BenchGen-Tasks}{746}
\DefMacro{BenchGen-TestsPerTask}{2.63}
\DefMacro{BenchGen-CodePerTask}{2.32}
\DefMacro{BenchGen-TestTokensPerTask}{344}
\DefMacro{BenchGen-CodeTokensPerTask}{316}
\DefMacro{BenchGen-TestExecutionTime}{59m 55s}
\DefMacro{BenchGen-AgentRunTime}{39h 2m 27s}
\DefMacro{BenchUpdate-Tasks}{509}
\DefMacro{BenchUpdate-TestsPerTask}{2.24}
\DefMacro{BenchUpdate-CodePerTask}{3.15}
\DefMacro{BenchUpdate-TestTokensPerTask}{989}
\DefMacro{BenchUpdate-CodeTokensPerTask}{424}
\DefMacro{BenchUpdate-TestExecutionTime}{43m 50s}
\DefMacro{BenchUpdate-AgentRunTime}{33h 21m 52s}

\DefMacro{MutGen-TotalTasks}{746}
\DefMacro{MutGen-TotalMethods}{1{,}961}
\DefMacro{MutGen-MethodsWithMutants}{1389.5}
\DefMacro{MutGen-MethodsWithMutantsPct}{70.9\%}
\DefMacro{MutGen-KilledPerMethod}{4.08}
\DefMacro{MutGen-KilledPerMethodPct}{56.1\%}
\DefMacro{MutGen-GeneratedPerMethod}{8.18}
\DefMacro{MutGen-NumAgentRuns}{4}
\DefMacro{MutUpdate-TotalTasks}{509}
\DefMacro{MutUpdate-TotalMethods}{1{,}138}
\DefMacro{MutUpdate-MethodsWithMutants}{554.8}
\DefMacro{MutUpdate-MethodsWithMutantsPct}{48.7\%}
\DefMacro{MutUpdate-KilledPerMethod}{3.85}
\DefMacro{MutUpdate-KilledPerMethodPct}{45.1\%}
\DefMacro{MutUpdate-GeneratedPerMethod}{8.70}
\DefMacro{MutUpdate-NumAgentRuns}{4}

\DefMacro{Gen-ClaudeCode-Success}{77.5\%}
\DefMacro{Gen-ClaudeCode-Redundant}{19.9\%}
\DefMacro{Gen-ClaudeCode-ExecFail}{0.3\%}
\DefMacro{Gen-ClaudeCode-CmplFail}{2.3\%}
\DefMacro{Gen-ClaudeCode-HrnsFail}{0.0\%}
\DefMacro{Gen-ClaudeCode-Cov}{74.8\%}
\DefMacro{Gen-ClaudeCode-CovOnPass}{76.8\%}
\DefMacro{Gen-ClaudeCode-CovOnPassHuman}{80.8\%}
\DefMacro{Gen-ClaudeCode-DeltaCovOnPass}{-4.043\%}
\DefMacro{Gen-ClaudeCode-Mut}{40.6\%}
\DefMacro{Gen-ClaudeCode-MutOnPass}{56.6\%}
\DefMacro{Gen-ClaudeCode-MutOnPassHuman}{54.2\%}
\DefMacro{Gen-ClaudeCode-DeltaMutOnPass}{+2.338\%}
\DefMacro{Gen-GeminiCLI-Success}{77.5\%}
\DefMacro{Gen-GeminiCLI-Redundant}{19.9\%}
\DefMacro{Gen-GeminiCLI-ExecFail}{0.4\%}
\DefMacro{Gen-GeminiCLI-CmplFail}{2.2\%}
\DefMacro{Gen-GeminiCLI-HrnsFail}{0.0\%}
\DefMacro{Gen-GeminiCLI-Cov}{73.2\%}
\DefMacro{Gen-GeminiCLI-CovOnPass}{75.1\%}
\DefMacro{Gen-GeminiCLI-CovOnPassHuman}{80.8\%}
\DefMacro{Gen-GeminiCLI-DeltaCovOnPass}{-5.699\%}
\DefMacro{Gen-GeminiCLI-Mut}{39.3\%}
\DefMacro{Gen-GeminiCLI-MutOnPass}{55.0\%}
\DefMacro{Gen-GeminiCLI-MutOnPassHuman}{54.2\%}
\DefMacro{Gen-GeminiCLI-DeltaMutOnPass}{+0.800\%}
\DefMacro{Gen-SWEAgentClaude-Success}{66.1\%}
\DefMacro{Gen-SWEAgentClaude-Redundant}{17.4\%}
\DefMacro{Gen-SWEAgentClaude-ExecFail}{0.3\%}
\DefMacro{Gen-SWEAgentClaude-CmplFail}{2.1\%}
\DefMacro{Gen-SWEAgentClaude-HrnsFail}{14.1\%}
\DefMacro{Gen-SWEAgentClaude-Cov}{65.1\%}
\DefMacro{Gen-SWEAgentClaude-CovOnPass}{78.0\%}
\DefMacro{Gen-SWEAgentClaude-CovOnPassHuman}{80.5\%}
\DefMacro{Gen-SWEAgentClaude-DeltaCovOnPass}{-2.518\%}
\DefMacro{Gen-SWEAgentClaude-Mut}{34.9\%}
\DefMacro{Gen-SWEAgentClaude-MutOnPass}{57.1\%}
\DefMacro{Gen-SWEAgentClaude-MutOnPassHuman}{55.3\%}
\DefMacro{Gen-SWEAgentClaude-DeltaMutOnPass}{+1.791\%}
\DefMacro{Gen-SWEAgentGemini-Success}{68.6\%}
\DefMacro{Gen-SWEAgentGemini-Redundant}{19.2\%}
\DefMacro{Gen-SWEAgentGemini-ExecFail}{0.1\%}
\DefMacro{Gen-SWEAgentGemini-CmplFail}{2.0\%}
\DefMacro{Gen-SWEAgentGemini-HrnsFail}{10.1\%}
\DefMacro{Gen-SWEAgentGemini-Cov}{65.3\%}
\DefMacro{Gen-SWEAgentGemini-CovOnPass}{74.3\%}
\DefMacro{Gen-SWEAgentGemini-CovOnPassHuman}{80.4\%}
\DefMacro{Gen-SWEAgentGemini-DeltaCovOnPass}{-6.100\%}
\DefMacro{Gen-SWEAgentGemini-Mut}{35.8\%}
\DefMacro{Gen-SWEAgentGemini-MutOnPass}{55.6\%}
\DefMacro{Gen-SWEAgentGemini-MutOnPassHuman}{54.4\%}
\DefMacro{Gen-SWEAgentGemini-DeltaMutOnPass}{+1.110\%}
\DefMacro{Gen-Best-Success}{77.5\%}
\DefMacro{Gen-Best-Success-Harness}{\ClaudeCode}
\DefMacro{Gen-Best-Success-Model}{\ClaudeOpus}
\DefMacro{Gen-Worst-Success}{66.1\%}
\DefMacro{Gen-Worst-Success-Harness}{\SWEAgent}
\DefMacro{Gen-Worst-Success-Model}{\ClaudeOpus}
\DefMacro{Update-ClaudeCode-Success}{74.4\%}
\DefMacro{Update-ClaudeCode-ExecFail}{23.8\%}
\DefMacro{Update-ClaudeCode-CmplFail}{1.8\%}
\DefMacro{Update-ClaudeCode-HrnsFail}{0.0\%}
\DefMacro{Update-ClaudeCode-Cov}{59.1\%}
\DefMacro{Update-ClaudeCode-CovOnPass}{79.4\%}
\DefMacro{Update-ClaudeCode-CovOnPassHuman}{80.0\%}
\DefMacro{Update-ClaudeCode-DeltaCovOnPass}{-0.585\%}
\DefMacro{Update-ClaudeCode-Mut}{19.2\%}
\DefMacro{Update-ClaudeCode-MutOnPass}{44.6\%}
\DefMacro{Update-ClaudeCode-MutOnPassHuman}{46.2\%}
\DefMacro{Update-ClaudeCode-DeltaMutOnPass}{-1.614\%}
\DefMacro{Update-GeminiCLI-Success}{74.6\%}
\DefMacro{Update-GeminiCLI-ExecFail}{23.6\%}
\DefMacro{Update-GeminiCLI-CmplFail}{1.4\%}
\DefMacro{Update-GeminiCLI-HrnsFail}{0.4\%}
\DefMacro{Update-GeminiCLI-Cov}{59.1\%}
\DefMacro{Update-GeminiCLI-CovOnPass}{79.1\%}
\DefMacro{Update-GeminiCLI-CovOnPassHuman}{80.0\%}
\DefMacro{Update-GeminiCLI-DeltaCovOnPass}{-0.876\%}
\DefMacro{Update-GeminiCLI-Mut}{19.4\%}
\DefMacro{Update-GeminiCLI-MutOnPass}{44.9\%}
\DefMacro{Update-GeminiCLI-MutOnPassHuman}{46.4\%}
\DefMacro{Update-GeminiCLI-DeltaMutOnPass}{-1.448\%}
\DefMacro{Update-SWEAgentClaude-Success}{65.6\%}
\DefMacro{Update-SWEAgentClaude-ExecFail}{19.1\%}
\DefMacro{Update-SWEAgentClaude-CmplFail}{4.3\%}
\DefMacro{Update-SWEAgentClaude-HrnsFail}{11.0\%}
\DefMacro{Update-SWEAgentClaude-Cov}{52.0\%}
\DefMacro{Update-SWEAgentClaude-CovOnPass}{79.2\%}
\DefMacro{Update-SWEAgentClaude-CovOnPassHuman}{80.5\%}
\DefMacro{Update-SWEAgentClaude-DeltaCovOnPass}{-1.257\%}
\DefMacro{Update-SWEAgentClaude-Mut}{16.8\%}
\DefMacro{Update-SWEAgentClaude-MutOnPass}{46.0\%}
\DefMacro{Update-SWEAgentClaude-MutOnPassHuman}{48.0\%}
\DefMacro{Update-SWEAgentClaude-DeltaMutOnPass}{-2.015\%}
\DefMacro{Update-SWEAgentGemini-Success}{73.9\%}
\DefMacro{Update-SWEAgentGemini-ExecFail}{19.8\%}
\DefMacro{Update-SWEAgentGemini-CmplFail}{2.8\%}
\DefMacro{Update-SWEAgentGemini-HrnsFail}{3.5\%}
\DefMacro{Update-SWEAgentGemini-Cov}{58.4\%}
\DefMacro{Update-SWEAgentGemini-CovOnPass}{79.1\%}
\DefMacro{Update-SWEAgentGemini-CovOnPassHuman}{79.9\%}
\DefMacro{Update-SWEAgentGemini-DeltaCovOnPass}{-0.846\%}
\DefMacro{Update-SWEAgentGemini-Mut}{19.1\%}
\DefMacro{Update-SWEAgentGemini-MutOnPass}{44.7\%}
\DefMacro{Update-SWEAgentGemini-MutOnPassHuman}{46.3\%}
\DefMacro{Update-SWEAgentGemini-DeltaMutOnPass}{-1.620\%}
\DefMacro{Update-Best-Success}{74.6\%}
\DefMacro{Update-Best-Success-Harness}{\GeminiCLI}
\DefMacro{Update-Best-Success-Model}{\GeminiPro}
\DefMacro{Update-Worst-Success}{65.6\%}
\DefMacro{Update-Worst-Success-Harness}{\SWEAgent}
\DefMacro{Update-Worst-Success-Model}{\ClaudeOpus}

\setcitestyle{authoryear,round}
\let\cite\citep

\title{\Title}

\author{%
\textbf{Jiale Amber Wang$^{1}$\quad Kaiyuan Wang$^{2}$\quad Pengyu Nie$^{1}$} \\[5pt]
\normalfont $^{1}$University of Waterloo\qquad $^{2}$Google\\[4pt]
\normalfont\small\texttt{jiale.wang@uwaterloo.ca, kaiyuanw@google.com, pynie@uwaterloo.ca}
}

\begin{document}
\maketitle

\begin{abstract}

Software tests and code evolve together: a code change should be followed by new or updated tests that record the new software behavior.
Yet existing test generation and update benchmarks often isolate the test from the code change, and rely on static metadata that does not verify whether a test is executable or semantically tied to the code change.
This makes it difficult to evaluate whether a test automation agent understands how a code change should propagate into the test suite.

We introduce \Bench, a benchmark of test and code co-evolution tasks mined from software repositories, with two tracks:
in \emph{test generation}, the agent shall write new tests to capture the new software behavior;
in \emph{test update}, the agent shall adapt failing existing tests to the changed software behavior.
Each task is anchored to a real commit history and packaged with environment configuration to support \emph{execution}-grounded metrics such as pass rate, coverage, and mutation score.
\Bench is also a \emph{live} benchmark: each task records the timestamp of the test and code changes, and new tasks are periodically mined by our automated pipeline, so evaluation can be restricted to tasks postdating a model's training cutoff to reduce data leakage risk.
The current snapshot contains \UseMacro{BenchGen-Tasks}\ test generation and \UseMacro{BenchUpdate-Tasks}\ test update tasks, curated from \NumCandidateRecords\ candidate co-evolution records across \NumCandidateProjects\ open-source Java projects.
We experiment with four state-of-the-art agents that combine strong harnesses (\ClaudeCode, \GeminiCLI, and \SWEAgent) with strong foundation models (\ClaudeOpus and \GeminiPro).
Results show that they achieve up to \UseMacro{Gen-Best-Success}\ success rate on test generation and \UseMacro{Update-Best-Success}\ on test update.
However, success rate is materially lower on the most recent benchmark tasks and drops significantly under limited per-task cost.
\footnote{\Bench leaderboard and data explorer are hosted at \url{\URL}.}

\end{abstract}

\section{Introduction}
\label{sec:intro}

Regression tests encode expectations about software behavior.
When code changes, developers often update existing tests or write additional tests in the same or subsequent revision to record the new software behavior.
Prior studies of test and code co-evolution show that this synchronization is central to software development and maintenance~\cite{ZaidmanETAL11CoEvolution,LubsenETAL09AssociationRules,MarsavinaETAL14FineGrainedCoEvolution,LevinYehudai17TestMaintenanceCoEvolution,LeDilavrecETAL21UntanglingCoEvolutions}.\footnote{We use the general term ``code'' to denote the non-testing part of the codebase, which some studies call ``production code.''}
A test edit is therefore not merely another code change, but evidence of how developers believed the code edit should be exercised.

Current benchmarks and evaluation frameworks do not capture this signal well.
Test generation benchmarks usually ask a system to synthesize tests for a fixed code snapshot rather than for a code change~\cite{FraserArcuri11EvoSuite,PachecoETAL07Randoop,LopsETAL25AgoneTest}.
Prior work on test update addresses one important part of this problem: adapting existing tests after code changes~\cite{HuETAL23CEPROT,YaraghiETAL25TARGET,LiuETAL24SYNTER,TestUpdaterETAL25}; however, it does not cover additional tests added to the test suite, which are another common way developers record changed behavior.
Across both settings, many datasets rely on static dependency analysis and diff signals as labels, without preserving enough execution context to rebuild the project, run the affected tests, or verify that an updated or generated test actually reflects the code change.

Recent large language models and agents increasingly target software engineering as an application domain and an arena for demonstrating their reasoning capabilities.
Harnesses from both industrial generative AI providers (e.g., \ClaudeCode~\cite{Anthropic25ClaudeCode}, \GeminiCLI~\cite{Google25GeminiCLI}, \OpenAICodexCLI~\cite{OpenAI25CodexCLI}, \Cursor~\cite{Anysphere23Cursor}) and academic research (e.g., \SWEAgent~\cite{YangETAL24SWEAgent}) can navigate repositories, run commands, and iteratively edit files, with the goal of completing complex software engineering tasks.
These capabilities are backed by foundational models that are trained on coding and reasoning capabilities (e.g., \ClaudeOpus~\cite{Anthropic26ClaudeOpus}, \GeminiPro~\cite{GoogleDeepMind25GeminiPro}, \GPTCodex~\cite{OpenAI25GPTCodex}, \CursorComposer~\cite{Anysphere26Composer}).
An executable benchmark is urgently needed to evaluate the true capabilities of these agents in maintaining and generating tests.

We introduce \Bench, an executable and live benchmark for test and code co-evolution mined from open-source and license-permissive software repositories' commit histories.
The benchmark comprises two tracks aligned with the two common ways developers maintain their test suites:
in \emph{test generation}, the agent shall write a new test that captures the new software behavior introduced by a code change;
in \emph{test update}, the agent shall edit a failing existing test to adapt it to the new software behavior.
In both tracks, \Bench evaluates the produced test by executing it, measuring execution-grounded metrics including success rate (defined as passing on and only on the new code version), pass rate, coverage, and mutation score.

Data quality is a top priority in \Bench.
We use static commit-history signals only to identify candidate test-code change pairs, and design a conservative pipeline that retains only those pairs where the test can be executed and is semantically tied to the code change.
Our benchmark construction pipeline consists of three phases:
(1)~mining candidate test-code change pairs by walking adjacent commits in Java Maven repositories that pass compilation and execution checks, using lightweight dynamic instrumentation to accurately collect test-code dependencies;
(2)~cleaning pairs with weak semantic relationships between test and code changes, or where the test execution cannot be reliably replicated;
(3)~performing cross-revision execution checks to label the task type and constructing the validated records into benchmark tasks.
The resulting tasks are high-quality ones where the developer's test generation/update intent is clear and the execution environment is available to evaluate agent-produced tests.

We open-source \Bench and its evaluation framework, and welcome community submissions to the leaderboard.
We initialize the leaderboard with four state-of-the-art harness--model configurations: \ClaudeCode and \GeminiCLI evaluated as complete systems, and \SWEAgent paired with \ClaudeOpus and with \GeminiPro.
Results show that the strongest configurations achieve up to \UseMacro{Gen-Best-Success}\ success rate on test generation and up to \UseMacro{Update-Best-Success}\ success rate on test update.
However, the success rate is materially lower on the most recent benchmark tasks and drops significantly when the per-task cost is capped.
Going forward, we will periodically extend \Bench with newly mined tasks, refresh the leaderboard with additional open-source and cost-efficient harness--model configurations, and publish dated snapshots so that the live benchmark continues to track progress while supporting contamination-aware evaluation.

This paper makes the following contributions:

\begin{itemize}[topsep=3pt,itemsep=1ex,partopsep=0ex,parsep=0ex,leftmargin=*]
\item \MyPara{Executable and Live Benchmark} We release \Bench, a benchmark of test and code co-evolution tasks with two tracks, test generation and test update. To our knowledge, \Bench is the first live benchmark and the largest executable benchmark for test and code co-evolution.
\item \MyPara{Benchmark Construction Pipeline} \Bench is constructed through a three-phase mining, cleaning, and packaging pipeline that grounds task labels in cross-revision test execution outcomes rather than static diff heuristics.
\item \MyPara{Evaluation Setup} To comprehensively evaluate an agent's test evolution capabilities, \Bench provides a reproducible execution environment that supports compilation, execution, coverage, and mutation-score metrics.
\item \MyPara{Experiments} We evaluate four state-of-the-art agents: while overall success rate is promising, challenges remain on the most recent benchmark tasks and under limited per-task cost.
\end{itemize}

\section{Related Work}
\label{sec:related}

\MyPara{Test and Code Co-evolution}
\citet{ZaidmanETAL11CoEvolution} introduced repository-mining views relating test-writing activity to coverage; \citet{LubsenETAL09AssociationRules} quantified co-evolution with association rules over code and test class changes; and \citet{MarsavinaETAL14FineGrainedCoEvolution} mined fine-grained co-evolution patterns in open-source systems.
\citet{LevinYehudai17TestMaintenanceCoEvolution} report that code fixes often land without complementary test maintenance, and follow-up work detects or predicts when tests should co-evolve with code changes~\cite{WangETAL21FacilitatingCoEvolution,LeDilavrecETAL21UntanglingCoEvolutions}.
\citet{SunETAL23RevisitingCoEvolution} further show that co-change heuristics used to mine such samples introduce substantial noise, motivating more careful identification.
\Bench builds on this line of work by turning co-evolution observations into executable benchmark tasks: commit co-changes are treated as candidate signals, and a task is retained only after build, test, dependency, and coverage validation.

\MyPara{Automated Test Generation}
Symbolic and search-based tools such as EvoSuite~\cite{FraserArcuri11EvoSuite} and Randoop~\cite{PachecoETAL07Randoop} generate large amounts of tests for a fixed code snapshot, maximizing coverage and/or mutation score.
LLM-based test generators include TestPilot~\cite{SchaeferETAL24TestPilot}, which generates JavaScript tests by prompting and refining on test failures, and \AgoneTest~\cite{LopsETAL25AgoneTest}, an automated Java framework that scores LLM-generated unit tests with project setup, context extraction, execution, coverage, mutation, and test-smell metrics; HumanEvalPack~\cite{MuennighoffETAL24OctoPack} adds test-related synthesis and repair tasks atop HumanEval.
\Bench differs by tying generation to a code-revision pair: a generated test must pass on the new revision and is separately classified by whether it captures the introduced behavior, connecting test generation to code evolution.

\MyPara{Automated Test Repair and Update}
Early research on test repair focused on rule-based approaches: ReAssert~\cite{DanielETAL09ReAssert} suggests repairs from a library of assertion-edit strategies, and \citet{MirzaaghaeiETAL14TestEvolution} propose heuristic algorithms for repairing tests when method declarations change.
LLM-based approaches are arising: CEPROT~\cite{HuETAL23CEPROT} frames test co-evolution as a two-stage learning problem fine-tuned on CodeT5~\cite{WangETAL21CodeT5}; TARGET~\cite{YaraghiETAL25TARGET}, SYNTER~\cite{LiuETAL24SYNTER}, \TestUpdater~\cite{TestUpdaterETAL25}, and ReAccept~\cite{ChiETAL25ReAccept} extend this with richer execution context (error messages, diffs, coverage); UTFix~\cite{RahmanETAL25UTFix} targets Python.
\Bench creates a large-scale executable benchmark that covers both newly inserted tests and existing-test updates, different from prior work that only uses static diff signals as labels.

\MyPara{Software Engineering Benchmarks}
SWE-bench~\cite{JimenezETAL24SWEBench} collects GitHub issues and packages them into software repair tasks with evaluation setup of test execution signals, and the human-validated SWE-bench Verified~\cite{ChowdhuryETAL24SWEBenchVerified} subset addresses underspecified issues and unreliable test cases in the original release.
SWT-Bench~\cite{MundlerETAL24SWTBench} reuses SWE-bench issues and asks code agents to generate tests that reproduce each bug-fix, scoring a test as valid when it fails before the fix and passes after; \Bench is closest in spirit but anchors tasks to commit-level code changes rather than issue narratives, and reports both update and generation alongside coverage and mutation signals.
LiveCodeBench~\cite{JainETAL24LiveCodeBench} introduces a timestamp-anchored, contamination-aware code generation benchmark that continuously ingests competitive-programming problems; \Bench shares this live-ingestion philosophy but applies it to executable code/test co-evolution tasks, supporting date-filtered evaluation over target revisions for contamination-aware reporting against any chosen training-cutoff window.

\section{Benchmark Construction}
\label{sec:benchmark}

\begin{figure}[t]
\centering
\includegraphics[width=.9\linewidth]{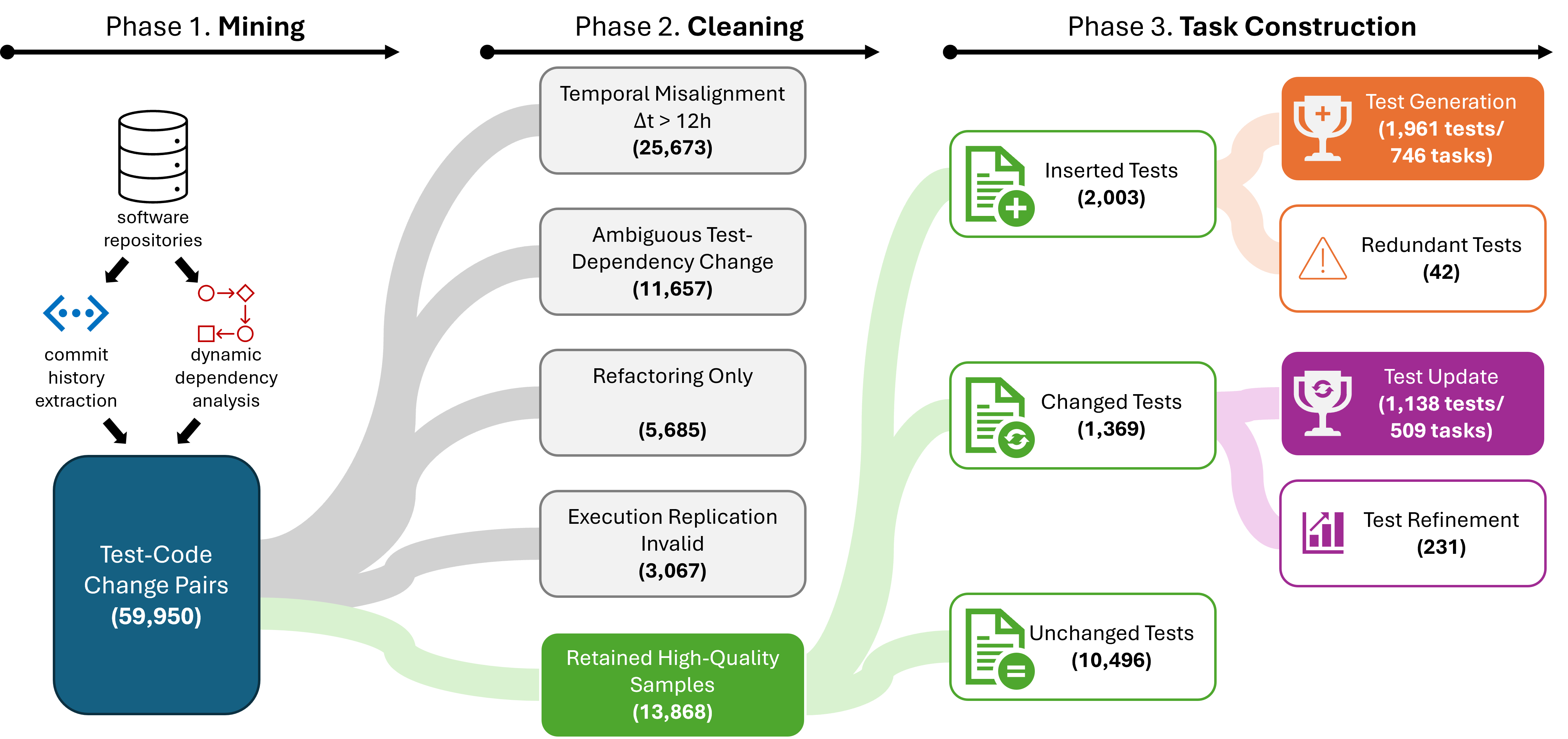}
\caption{\Bench is constructed through a three-phase mining, cleaning, and task construction pipeline.}
\label{fig:data-pipeline}
\end{figure}

\Bench is constructed by a fully automated pipeline (Figure~\ref{fig:data-pipeline}) with three phases: mining test-code change pairs (\S\ref{sec:benchmark:mining}), execution-based data cleaning (\S\ref{sec:benchmark:cleaning}), and benchmark task construction (\S\ref{sec:benchmark:task-construction}).
\S\ref{sec:benchmark:statistics} reports benchmark statistics.

\subsection{Phase 1: Mining Test-Code Change Pairs}
\label{sec:benchmark:mining}

The first phase identifies candidate test-code change pairs by mining open-source and license-permissive repositories.
We focus on repositories that use Java programming language and Maven build system because of the their well-established building and testing conventions, which helps improve the quality of the collected data.
\Bench discovers candidate Java Maven repositories through the GitHub Search API, requires the presence of a top-level \texttt{pom.xml}, and filters by license, star count, clone success, Java version, and baseline build success on the latest (release) commit.
For each retained repository, \Bench walks the most recent commits before a fixed date cutoff and forms adjacent commit pairs that build cleanly and pass tests on both endpoints.
For every surviving commit pair, \Bench extracts method-level code and test diffs and runs static and runtime analysis to recover method metadata, source locations, classpaths, module information, and per-test runtime dependencies on production methods.
Repository-level and commit-level details are documented in Appendix~\ref{sec:appendix:details:mining}.

\subsection{Phase 2: Execution-Backed Data Cleaning}
\label{sec:benchmark:cleaning}

The second phase rebuilds candidate revision pairs and executes the relevant tests across both old and new revisions.
These runs serve two purposes.
First, they ensure that a task can be evaluated automatically rather than merely appearing in a diff.
Second, they provide the cross-revision outcomes needed to label task type and to compute the discriminating signal during evaluation.
\Bench executes the four cross-revision triples \OldOnOld, \NewOnNew, \OldOnNew, and \NewOnOld where applicable: an update task requires that the old test passes on the old code and the new test passes on the new code, while the old test fails or fails to compile on the new code; a generation task requires that the new test passes on the new code, and is further classified as discriminating when the new test also fails on the old code.
\Bench then applies aggressive filters: every retained test-code change pair must have a passing new-revision test, no dependencies on concurrent changes in the test directory, non-refactoring test and code changes (refactorings are detected with RefactoringMiner~\cite{TsantalisETAL20RefactoringMiner}), and a revision-pair gap of at most 12 hours (following prior co-evolution-mining practice~\cite{SunETAL23RevisitingCoEvolution}).
These data cleaning steps intentionally discard many plausible-looking examples, because a benchmark task is only useful if it can be rebuilt, executed, and scored without hidden manual judgment.

\subsection{Phase 3: Task Construction}
\label{sec:benchmark:task-construction}

The third phase converts validated test-code change pairs into benchmark tasks.
Among the inserted tests, \Bench further filters out \NewOnOld pass cases (i.e., the new test is passing on the old code), because they do not semantically depend on the current code change thus are considered as ``redundant''.
Among the changed tests, \Bench further filters out \OldOnOld pass cases: these tests are being \emph{refined} rather than updated; since test refinement can be triggered by various reasons from typos to coverage improvements~\cite{SunETAL23RevisitingCoEvolution}, we choose to not include them in the benchmark and focus on the explicit functional update cases.
We also collected tests are unchanged (but whose dependency code changed) and use them as negative samples of the test update identification task, studied in Appendix~\ref{sec:appendix:test-update-identification}.
For test generation and test update, we group the tests that are changed in the same commit pair into a single task, because they share the same software behavior change and should be addressed in one agent run in practice.
Each task includes the old and new tests (only new tests for test generation), changed code dependencies, revision metadata, and execution signals (pass/fail, coverage, and mutation score) of the developer-written tests.

\subsection{Benchmark Statistics}
\label{sec:benchmark:statistics}

\begin{table*}[t]
\caption{\Bench statistics.
Token counts are obtained using Gemini's tokenizer.
Test exec time is the wall-clock time to run all tests under the track once.
Agent run time is the wall-clock time to run an agent to solve all tasks, including LLM calls, tool calls, and test execution, averaged across the four agents experimented in this paper.}
\label{tab:benchmark-track-statistics}
\centering
\small
\resizebox{\linewidth}{!}{%

\begin{tabular}{lrrrrrrr}
\toprule
\textbf{Track} & \textbf{\#Task} & \textbf{\#Test/Task} & \textbf{\#Code/Task} & \textbf{\#Test Token/Task} & \textbf{\#Code Token/Task} & \textbf{Test Exec Time} & \textbf{Agent Run Time} \\
\midrule
Test generation & \UseMacro{BenchGen-Tasks} & \UseMacro{BenchGen-TestsPerTask} & \UseMacro{BenchGen-CodePerTask} & \UseMacro{BenchGen-TestTokensPerTask} & \UseMacro{BenchGen-CodeTokensPerTask} & \UseMacro{BenchGen-TestExecutionTime} & \UseMacro{BenchGen-AgentRunTime} \\
Test update & \UseMacro{BenchUpdate-Tasks} & \UseMacro{BenchUpdate-TestsPerTask} & \UseMacro{BenchUpdate-CodePerTask} & \UseMacro{BenchUpdate-TestTokensPerTask} & \UseMacro{BenchUpdate-CodeTokensPerTask} & \UseMacro{BenchUpdate-TestExecutionTime} & \UseMacro{BenchUpdate-AgentRunTime} \\
\bottomrule
\end{tabular}

}
\end{table*}

\Bench is constructed with a data cutoff at the end of 2025.
The current task-construction run starts from \NumCandidateRecords\ candidate test-change records sourced from \NumCandidateProjects\ post-mining repositories that pass the Phase~1--2 license, star, build, JUnit, and commit-level filters, and retains \NumClassifiedRecords\ classified records.
Table~\ref{tab:benchmark-track-statistics} reports the resulting statistics:
\Bench contains \UseMacro{BenchGen-Tasks}\ test-generation tasks (averaging \UseMacro{BenchGen-TestsPerTask}\ tests and \UseMacro{BenchGen-CodePerTask}\ dependency code methods per task) and \UseMacro{BenchUpdate-Tasks}\ test-update tasks (averaging \UseMacro{BenchUpdate-TestsPerTask}\ tests and \UseMacro{BenchUpdate-CodePerTask}\ dependency code methods per task).
The task input/output size is moderate (hundreds of tokens) but each task requires non-trivial execution, reasoning, and refinement to complete:
a typical evaluation run of an agent on all \Bench tasks takes about 72 machine-hours, parallelizable across machines.

\begin{figure}[t]
\centering
\setlength{\abovecaptionskip}{2pt}
\setlength{\belowcaptionskip}{0pt}
\includegraphics[width=.9\linewidth]{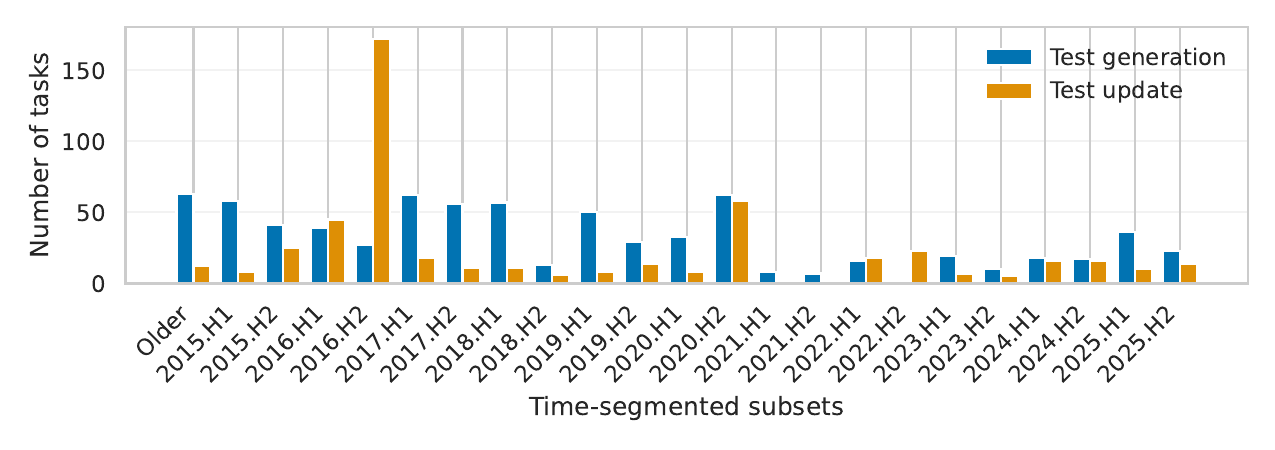}
\vspace{-0.5em}
\caption{\Bench task counts across time-segmented subsets. Each bar reports the number of tasks whose new revision falls in the corresponding half-year; the \emph{Older} subset combines all tasks before 2015.}
\label{fig:benchmark-tasks-by-time}
\vspace{-1em}
\end{figure}

\MyPara{Live Benchmark}
Because each task is anchored to a concrete timestamp, \Bench naturally decomposes into time-segmented subsets.
This supports contamination-aware evaluation: a leaderboard can restrict scoring to tasks authored after a model's training cutoff.
Figure~\ref{fig:benchmark-tasks-by-time} reports the temporal distribution of the tasks in the two tracks. %
The benchmark website exposes these subsets through a data explorer and leaderboard. %

\MyPara{Data Availability}
\Bench is released at \url{https://huggingface.co/TestEvo-Bench/datasets} under the CC BY 4.0 license. %

\section{Benchmark Evaluation Setup}
\label{sec:setup}

\subsection{Task Input and Output Format}
\label{sec:setup:tasks}

Each \Bench task is specified at the level of test changes in a single commit pair.
A task provides the repository URL, old and new revisions, build system and execution environment configuration (e.g., classpath for Java), the relevant test file, the changed code methods, the focal code context, and the tests to be generated or updated.
An agent shall produce a patch with the generated or updated tests.
\Bench evaluation framework then applies the patch, compiles the project, runs the selected tests, and records the per-stage outcome (injection, compile, test) along with coverage and, when available, mutation score.
Below we summarize the instructions given to the agent in the two tracks; Appendix~\ref{sec:appendix:prompts} provides the full prompts.

\begin{figure}[t]
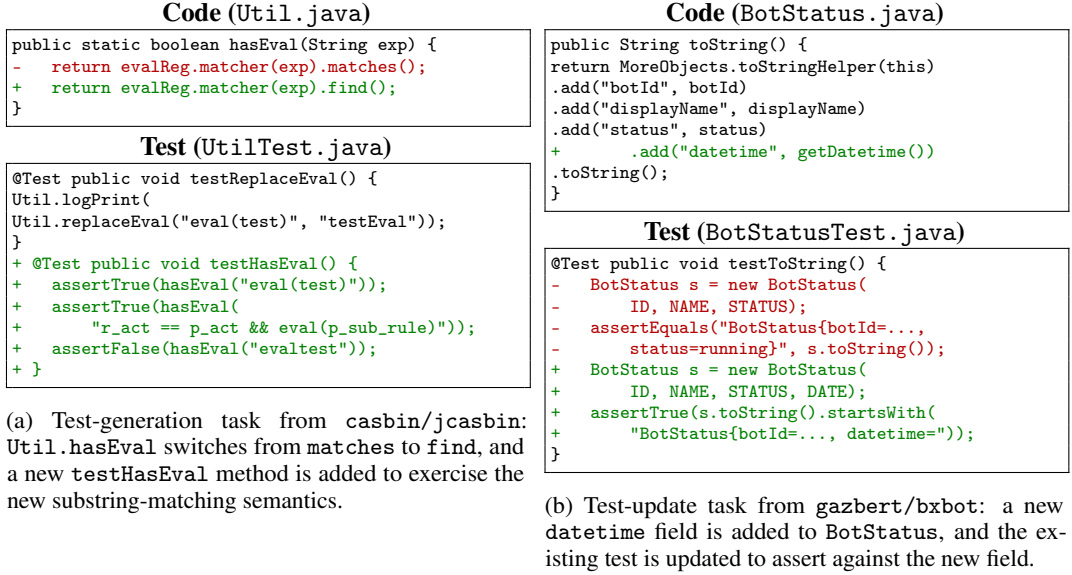

\centering
\begin{subfigure}[t]{0.49\linewidth}
\centering
\textbf{Code (\texttt{Util.java})}
\begin{lstlisting}[language=java-pretty,basicstyle=\scriptsize\ttfamily,frame=single,framesep=2pt,xleftmargin=2pt,numbers=none,escapeinside={(*@}{@*)},aboveskip=2pt,belowskip=2pt]
public static boolean hasEval(String exp) {
(*@\textcolor{red!70!black}{-~~~return evalReg.matcher(exp).matches();}@*)
(*@\textcolor{green!50!black}{+~~~return evalReg.matcher(exp).find();}@*)
}
\end{lstlisting}
\textbf{Test (\texttt{UtilTest.java})}
\begin{lstlisting}[language=java-pretty,basicstyle=\scriptsize\ttfamily,frame=single,framesep=2pt,xleftmargin=2pt,numbers=none,escapeinside={(*@}{@*)},aboveskip=2pt,belowskip=2pt]
@Test public void testReplaceEval() {
Util.logPrint(
Util.replaceEval("eval(test)", "testEval"));
}
(*@\textcolor{green!50!black}{+ @Test public void testHasEval() \{}@*)
(*@\textcolor{green!50!black}{+~~~assertTrue(hasEval("eval(test)"));}@*)
(*@\textcolor{green!50!black}{+~~~assertTrue(hasEval(}@*)
(*@\textcolor{green!50!black}{+~~~~~~~"r\_act == p\_act \&\& eval(p\_sub\_rule)"));}@*)
(*@\textcolor{green!50!black}{+~~~assertFalse(hasEval("evaltest"));}@*)
(*@\textcolor{green!50!black}{+ \}}@*)
\end{lstlisting}
\caption{Test-generation task from \texttt{casbin/jcasbin}: \texttt{Util.hasEval} switches from \texttt{matches} to \texttt{find}, and a new \texttt{testHasEval} method is added to exercise the new substring-matching semantics.}
\label{fig:task-examples:generation}
\end{subfigure}
\hfill
\begin{subfigure}[t]{0.49\linewidth}
\centering
\textbf{Code (\texttt{BotStatus.java})}
\begin{lstlisting}[language=java-pretty,basicstyle=\scriptsize\ttfamily,frame=single,framesep=2pt,xleftmargin=2pt,numbers=none,escapeinside={(*@}{@*)},aboveskip=2pt,belowskip=2pt]
public String toString() {
return MoreObjects.toStringHelper(this)
.add("botId", botId)
.add("displayName", displayName)
.add("status", status)
(*@\textcolor{green!50!black}{+~~~~~~~.add("datetime", getDatetime())}@*)
.toString();
}
\end{lstlisting}
\textbf{Test (\texttt{BotStatusTest.java})}
\begin{lstlisting}[language=java-pretty,basicstyle=\scriptsize\ttfamily,frame=single,framesep=2pt,xleftmargin=2pt,numbers=none,escapeinside={(*@}{@*)},aboveskip=2pt,belowskip=2pt]
@Test public void testToString() {
(*@\textcolor{red!70!black}{-~~~BotStatus s = new BotStatus(}@*)
(*@\textcolor{red!70!black}{-~~~~~~~ID, NAME, STATUS);}@*)
(*@\textcolor{red!70!black}{-~~~assertEquals("BotStatus\{botId=...,}@*)
(*@\textcolor{red!70!black}{-~~~~~~~status=running\}", s.toString());}@*)
(*@\textcolor{green!50!black}{+~~~BotStatus s = new BotStatus(}@*)
(*@\textcolor{green!50!black}{+~~~~~~~ID, NAME, STATUS, DATE);}@*)
(*@\textcolor{green!50!black}{+~~~assertTrue(s.toString().startsWith(}@*)
(*@\textcolor{green!50!black}{+~~~~~~~"BotStatus\{botId=..., datetime="));}@*)
}
\end{lstlisting}
\caption{Test-update task from \texttt{gazbert/bxbot}: a new \texttt{datetime} field is added to \texttt{BotStatus}, and the existing test is updated to assert against the new field.}
\label{fig:task-examples:update}
\end{subfigure}
\caption{Representative tasks from the two \Bench tracks. Each task ships the (old, new) revision pair, the focal production method, the test method(s), and the metadata required by the runner. Each panel shows the production-code diff (top) and the corresponding test diff (bottom).}
\label{fig:task-examples}
\end{figure}

\MyPara{Test generation}
On the test generation track, the agent shall add tests that exercise the new software behavior introduced by the code changes.
A successful test must pass on the new revision and \emph{fail} on the old revision, which demonstrates that it captured the introduced behavior.
A test that passes on both revisions may improve coverage but does not certify that the agent perceived the behavioral difference.
Figure~\ref{fig:task-examples:generation} shows a representative generation task from \texttt{casbin/jcasbin}: \texttt{Util.hasEval} switches from \texttt{matches} to \texttt{find}, and a new \texttt{testHasEval} method is added to exercise the new substring-matching semantics.

\MyPara{Test update}
On the test update track, the starting repository is the new code revision with the target tests reverted to their old versions.
The agent shall edit the existing tests so they compile and pass on the new revision.
The developer-updated tests are not used as a textual oracle: an alternative test is acceptable as long as it integrates with the project, passes on the new revision, and exercises the changed behavior.
This framing rewards behavior-preserving updates rather than surface-level reproduction of the developer's edits.
Figure~\ref{fig:task-examples:update} shows a representative update task from \texttt{gazbert/bxbot}: the production class \texttt{BotStatus} gains a \texttt{datetime} field, and the existing test is updated to construct \texttt{BotStatus} with the new field and to assert against it.

\subsection{Execution Environment}
\label{sec:setup:env}

\Bench provides an execution environment shared across both tracks: it checks out the new revision, applies the candidate edit, recompiles only the affected module to localize compilation errors (using Maven, with handling for multi-module projects), and executes each target test in an isolated JVM (using JUnit).
Coverage is collected with JaCoCo~\cite{JaCoCo} and intersected with the focal dependency lines recorded at task-construction time; mutation analysis uses Universal Mutator~\cite{GroceETAL18UniversalMutator} restricted to the focal dependency lines of the new revision.
The evaluation framework records per-method outcomes (pass, execution failure, compile failure, or harness/code-injection failure) along with stdout, stderr, and coverage and mutation analysis results.
Full configuration details are documented in Appendix~\ref{sec:appendix:details:environment}. %

\subsection{Metrics}
\label{sec:setup:metrics}

For each task we partition its target tests into mutually exclusive outcome categories and report the per-task rate of each category, averaged across tasks (macro average).
\SuccessPct\ is the primary success rate metric for both tracks: the agent-produced test must pass on the new revision and, on the test generation track, additionally fail on the old revision (either by an assertion failure or by a compile error).
On the test update track, the old test fails on the old code by design, thus \SuccessPct\ reduces to passing on the new revision.
The metrics indicating non-success include:
\RedundantPct\ (test generation only; passes on the new revision but does not fail on the old, so the test does not capture the introduced behavior),
\ExecFailPct\ (produced test failed on the new revision),
\CmplFailPct\ (produced test cannot compile),
and \HrnsFailPct\ (the harness produced no test or a malformed edit that cannot be applied).

We also measure the adequacy of the produced tests (i.e., whether they are sufficient to effectively assess the code change) using coverage and mutation score.
Coverage measures the portion of the code that is executed during the test.
Mutation score measures how many mutants are killed by the test, which is a proxy for the test's ability to detect bugs in the code.
We report the average coverage and mutation score only for the produced tests that passed on the new revision, denoted as \CovOnPassMetric\ and \MutOnPassMetric\ respectively, since the two metrics are not meaningful for failing tests.
Appendix~\ref{sec:appendix:additional-results} presents the raw coverage and mutation scores and comparisons with the developer-written tests.

\section{Experiments}
\label{sec:experiments}

\begin{wraptable}{r}{0.42\linewidth}
\vspace{-1.0em}
\centering
\caption{Harness--model configurations evaluated by \Bench.}
\label{tab:harnesses}
\small

\begin{tabular}{ll}
\toprule
\textbf{Harness} & \textbf{Model} \\
\midrule
\ClaudeCode & \ClaudeOpus \\
\midrule
\GeminiCLI  & \GeminiPro \\
\midrule
\multirow{2}{*}{\SWEAgent}   & \ClaudeOpus \\
& \GeminiPro \\
\bottomrule
\end{tabular}

\vspace{-0.8em}
\end{wraptable}%
We evaluate four harness--model configurations (agents) summarized in Table~\ref{tab:harnesses}. %
\ClaudeCode~\cite{Anthropic25ClaudeCode} and \GeminiCLI~\cite{Google25GeminiCLI}, which are state-of-the-art coding agent products from industrial generative AI providers, represent agents where the harness and foundation model are co-designed and tightly integrated.
\SWEAgent~\cite{YangETAL24SWEAgent} is a coding agent harness originating from academic research, and we pair it with the same foundation models as the industrial harnesses (i.e., \ClaudeOpus~\cite{Anthropic26ClaudeOpus} and \GeminiPro~\cite{GoogleDeepMind25GeminiPro}) to facilitate fair comparison.

\begin{table*}[t]
\centering
\caption{Experimental results of state-of-the-art agents on \Bench.
All metrics are per-task (macro) averages.
\SuccessPct is the percentage of tasks where the agent-produced tests correctly reflect the code change by passing on the new revision and failing on the old revision.
\RedundantPct, \ExecFailPct, \CmplFailPct, and \HrnsFailPct are failure modes (cf \S\ref{sec:setup:metrics}).
\CovOnPassMetric\ and \MutOnPassMetric\ are coverage and mutation score of generated passing tests.
Best metrics are bolded.}
\label{tab:main-results}
\begin{subtable}{\linewidth}
\centering
\subcaption{Test generation on \NumGenerationMethodsCurrent\ methods across \UseMacro{BenchGen-Tasks}\ tasks.}
\label{tab:test-generation-results}
\small
\resizebox{\linewidth}{!}{%

\begin{tabular}{llrrrrrrr}
\toprule
\textbf{\UseMacro{TH-Harness}} & \textbf{\UseMacro{TH-Model}} & \textbf{\UseMacro{TH-SuccessPct}\,$\uparrow$} & \textbf{\UseMacro{TH-RedundantPct}\,$\downarrow$} & \textbf{\UseMacro{TH-ExecFailPct}\,$\downarrow$} & \textbf{\UseMacro{TH-CmplFailPct}\,$\downarrow$} & \textbf{\UseMacro{TH-HrnsFailPct}\,$\downarrow$} & \textbf{\UseMacro{TH-CovOnPass}\,$\uparrow$} & \textbf{\UseMacro{TH-MutOnPass}\,$\uparrow$} \\
\midrule
\ClaudeCode & \ClaudeOpus & \textbf{77.5\%} & 19.9\% & 0.3\% & 2.3\% & \textbf{0.0\%} & 76.8\% & 56.6\% \\
\GeminiCLI & \GeminiPro & \textbf{77.5\%} & 19.9\% & 0.4\% & 2.2\% & \textbf{0.0\%} & 75.1\% & 55.0\% \\
\SWEAgent & \ClaudeOpus & 66.1\% & \textbf{17.4\%} & 0.3\% & 2.1\% & 14.1\% & \textbf{78.0\%} & \textbf{57.1\%} \\
\SWEAgent & \GeminiPro & 68.6\% & 19.2\% & \textbf{0.1\%} & \textbf{2.0\%} & 10.1\% & 74.3\% & 55.6\% \\
\bottomrule
\end{tabular}

}
\end{subtable}

\vspace{0.6em}

\begin{subtable}{\linewidth}
\centering
\subcaption{Test update on \NumUpdateMethodsCurrent\ methods across \UseMacro{BenchUpdate-Tasks}\ tasks.}
\label{tab:test-update-results}
\small
\resizebox{\linewidth}{!}{%

\begin{tabular}{llrrrrrr}
\toprule
\textbf{\UseMacro{TH-Harness}} & \textbf{\UseMacro{TH-Model}} & \textbf{\UseMacro{TH-SuccessPct}\,$\uparrow$} & \textbf{\UseMacro{TH-ExecFailPct}\,$\downarrow$} & \textbf{\UseMacro{TH-CmplFailPct}\,$\downarrow$} & \textbf{\UseMacro{TH-HrnsFailPct}\,$\downarrow$} & \textbf{\UseMacro{TH-CovOnPass}\,$\uparrow$} & \textbf{\UseMacro{TH-MutOnPass}\,$\uparrow$} \\
\midrule
\ClaudeCode & \ClaudeOpus & 74.4\% & 23.8\% & 1.8\% & \textbf{0.0\%} & \textbf{79.4\%} & 44.6\% \\
\GeminiCLI & \GeminiPro & \textbf{74.6\%} & 23.6\% & \textbf{1.4\%} & 0.4\% & 79.1\% & 44.9\% \\
\SWEAgent & \ClaudeOpus & 65.6\% & \textbf{19.1\%} & 4.3\% & 11.0\% & 79.2\% & \textbf{46.0\%} \\
\SWEAgent & \GeminiPro & 73.9\% & 19.8\% & 2.8\% & 3.5\% & 79.1\% & 44.7\% \\
\bottomrule
\end{tabular}

}
\end{subtable}
\end{table*}

\subsection{Test Generation Results}
\label{sec:experiments:generation}

Table~\ref{tab:test-generation-results} reports test-generation results.
\ClaudeCode and \GeminiCLI tie for the highest \SuccessPct\ at \UseMacro{Gen-ClaudeCode-Success}, followed by \SWEAgent + \GeminiPro\ at \UseMacro{Gen-SWEAgentGemini-Success}\ and \SWEAgent + \ClaudeOpus\ at \UseMacro{Gen-SWEAgentClaude-Success}, spanning more than ten percentage points across the four agents.
Notably, \RedundantPct\ remains sizeable (\UseMacro{Gen-SWEAgentClaude-Redundant}--\UseMacro{Gen-ClaudeCode-Redundant}) on every agent; this suggests that the agents are strong at generating passing tests, but sometimes struggle to create non-trivial test oracles that are semantically related to the code change.
Execution and compilation failures are small (\ExecFailPct\ at most \UseMacro{Gen-GeminiCLI-ExecFail}, \CmplFailPct\ at most \UseMacro{Gen-ClaudeCode-CmplFail}).
The gap on \SWEAgent\ comes from \HrnsFailPct\ (\UseMacro{Gen-SWEAgentClaude-HrnsFail}\ with \ClaudeOpus, \UseMacro{Gen-SWEAgentGemini-HrnsFail}\ with \GeminiPro), demonstrating some level of incompatibility between the harness and foundation models.
\CovOnPassMetric\ falls between \UseMacro{Gen-SWEAgentGemini-CovOnPass}\ and \UseMacro{Gen-SWEAgentClaude-CovOnPass}\ across agents, and \MutOnPassMetric\ between \UseMacro{Gen-GeminiCLI-MutOnPass}\ and \UseMacro{Gen-SWEAgentClaude-MutOnPass}.

\subsection{Test Update Results}
\label{sec:experiments:update}

Table~\ref{tab:test-update-results} reports test-update results.
\GeminiCLI\ (\UseMacro{Update-GeminiCLI-Success}), \ClaudeCode\ (\UseMacro{Update-ClaudeCode-Success}), and \SWEAgent + \GeminiPro\ (\UseMacro{Update-SWEAgentGemini-Success}) achieve nearly identical \SuccessPct, while \SWEAgent + \ClaudeOpus\ trails at \UseMacro{Update-SWEAgentClaude-Success}; the gap is dominated by \HrnsFailPct\ (\UseMacro{Update-SWEAgentClaude-HrnsFail}\ versus at most \UseMacro{Update-SWEAgentGemini-HrnsFail}\ on the other agents) rather than by execution or compilation failures.
\CovOnPassMetric\ is tightly clustered between \UseMacro{Update-GeminiCLI-CovOnPass}\ and \UseMacro{Update-ClaudeCode-CovOnPass}, and \MutOnPassMetric\ between \UseMacro{Update-ClaudeCode-MutOnPass}\ and \UseMacro{Update-SWEAgentClaude-MutOnPass}.

\subsection{Performance Over Time}
\label{sec:experiments:over-time}

\begin{figure}[t]
\centering
\includegraphics[width=\linewidth]{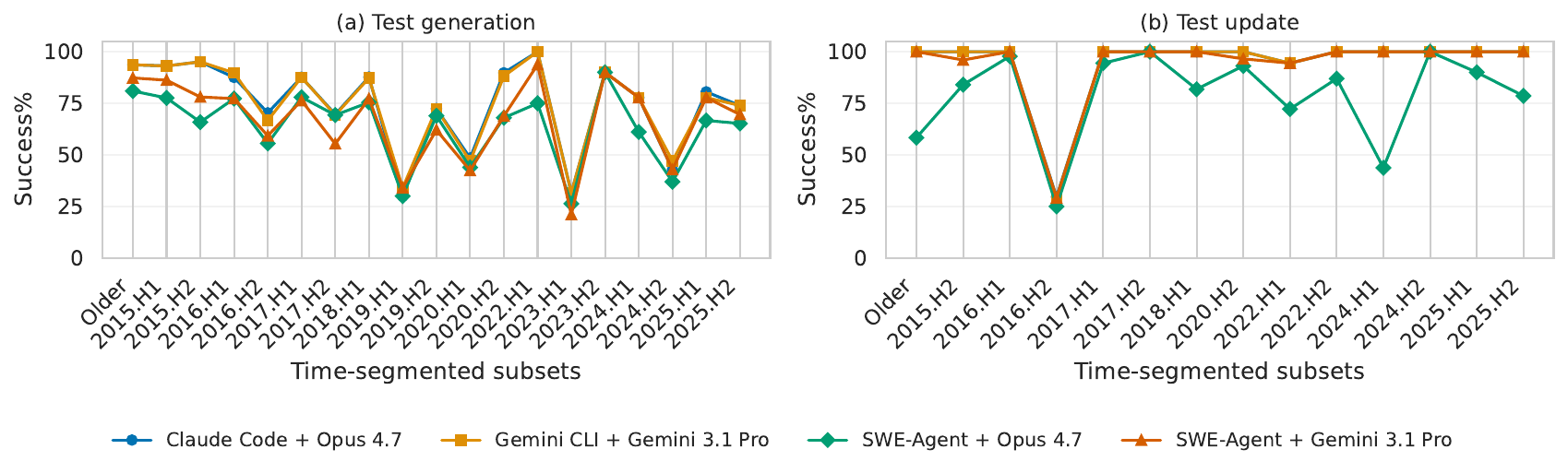}
\caption{\SuccessPct\ across time-segmented subsets for (a)~test generation and (b)~test update. Subsets with fewer than $20$ methods are omitted to suppress small-sample noise.}
\label{fig:perf-over-time}
\vspace{-2em}
\end{figure}

Figure~\ref{fig:perf-over-time} shows \SuccessPct\ by time-segmented subsets. %
For test generation,
we observe a gradual drop as the data gets more recent, which indicates that the agents that are predominantly trained on older data may not generalize well to newer data.
For test update, we observe a more stable performance across time-segments except for an outlier in 2016.H2 (caused by large number of failures in one repository).
We hypothesize that the update from old tests to new tests usually follows some patterns that are persistent across time, thus is less sensitive to temporal changes.

\subsection{Cost Analysis}
\label{sec:experiments:cost}

\begin{figure}[t]
\centering
\includegraphics[width=\linewidth]{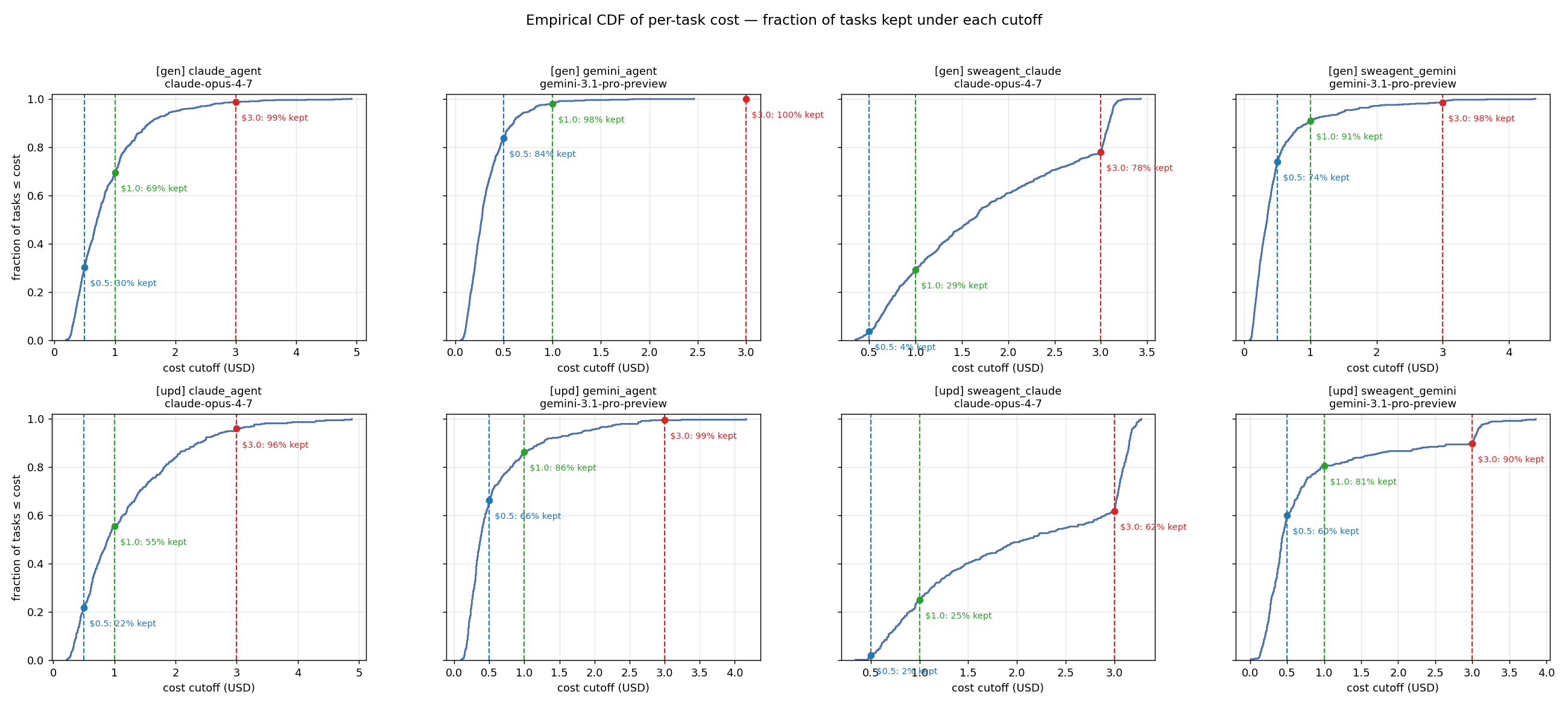}
\caption{Empirical CDF when per-task cost is limited. Top row: test generation; bottom row: test update. Vertical dashed lines mark per-task budget caps at \$3 (red), \$1 (green), and \$0.5 (blue); the annotated ``\% kept'' is the fraction of tasks whose default-budget cost stays at or below each cap.}
\label{fig:cost-ecdf}
\vspace{-2em}
\end{figure}

We run each agent with a per-task target budget of \$3 (the last round of LLM call may exceed the budget, as we can only monitor the cost and stop the agent after each round of LLM call).
We additionally study tighter operating points by capping the per-task spend at \$1 and \$0.5, treating tasks whose default-budget cost exceeds the cap as harness failures.
Figure~\ref{fig:cost-ecdf} reports the empirical CDF of per-task billed cost on both tracks; Table~\ref{tab:cost-cutoff-results} (Appendix~\ref{sec:appendix:additional-results}) reports \SuccessPct\ and \HrnsFailPct\ at each cap.
The two \ClaudeOpus-driven agents are cost-sensitive: at the \$1 cap, \ClaudeCode\ \SuccessPct\ drops from 70.6\% to 44.2\% on test generation and from 86.1\% to 54.8\% on test update, and \SWEAgent + \ClaudeOpus\ drops from 59.5\% to 18.8\% and from 73.2\% to 18.1\% respectively.
The two \GeminiPro-driven agents are markedly more robust, but still show a drop in \SuccessPct as the budget is tightened.
Arguably, test evolution is a routine task that needs to be done frequently, thus improving the cost efficiency of the agents is important for the practicality of using them.

\section{Limitations and Future Work}
\label{sec:discussion}

\MyPara{Programming language and build system}
We focus on Java Maven projects because of their well-established building and testing conventions.
The benchmark construction pipeline is agnostic to the programming language and build system in principle; only the dependency, coverage, and mutation analyses techniques need to be adapted.
We plan to extend \Bench to more programming languages and build systems, such as Python and JavaScript that have been studied in the testing literature.

\MyPara{Model and harness coverage}
The four agents experimented in our work span open- and closed-source harnesses but represent only a small subset of the possible configurations nowadays.
We welcome community submissions to our leaderboard, meanwhile we plan to include open-source models and alternative harnesses in the near future.

\MyPara{Cost of execution-based evaluation}
\Bench's evaluation setup, including compilation, test execution, coverage, and mutation analysis, is much more expensive than diff-similarity scoring but is what makes the benchmark reproducible and reliable.
In our experiments, we mitigate this cost through parallelizing the tasks over a multi-machine cluster infrastructure and caching data (such as mutants) to avoid redundant computation, effectively bringing down the 72h wall-clock time to a few hours if LLM calls are not rate-limited.
We plan to make our infrastructure accessible through a public API, to make it easier for the community to evaluate their own agents without configuring sophisticated infrastructure.

\section{Conclusions}
\label{sec:conclusion}

We presented \Bench, an executable and live benchmark for test and code co-evolution.
A three-phase pipeline of mining, execution-backed cleaning, and task construction turns repository history into \UseMacro{BenchGen-Tasks}\ test-generation tasks (\NumGenerationMethodsCurrent\ methods) and \UseMacro{BenchUpdate-Tasks}\ test-update tasks (\NumUpdateMethodsCurrent\ methods) drawn from open-source Java Maven repositories, each packaged with the environment required to reproduce its build, test, coverage, and mutation evaluation.
Anchoring every task to a concrete revision date supports contamination-aware reporting on time-segmented subsets, and continuous mining keeps the benchmark aligned with how software evolves.
Across four state-of-the-art agents, the strongest reach \UseMacro{Gen-Best-Success}\ success rate on test generation and \UseMacro{Update-Best-Success}\ on test update; both are materially lower on the most recent subsets and drop sharply under tight per-task cost caps, indicating that neither track is yet saturated.
We release \Bench, its evaluation runner, and an open leaderboard at \url{\URL}.

\begin{ack}
We thank Saarang Agarwal, Yuntian Deng, Austing Dong, Liliana Hotsko, Mohammad Jaffer Iqbal, Bihui Jin, Yinxi Li, Yu Liu, Chengnian Sun
and the anonymous reviewers for their comments and feedback.
This work is enabled in part by support provided by Compute Ontario (computeontario.ca) and the Digital Research Alliance of Canada (alliancecan.ca).
This work is partially supported by Dragon Testing.
This work is also partially supported by the Natural Sciences and Engineering Research Council of Canada (NSERC) under funding reference number RGPIN2024-04909.
\end{ack}

\bibliography{bib}

\clearpage\newpage
\appendix

\section{\Bench Technical Details}
\label{sec:appendix:details}

\subsection{Mining Software Repositories}
\label{sec:appendix:details:mining}

\begin{figure}[t]
\centering
\includegraphics[width=\linewidth]{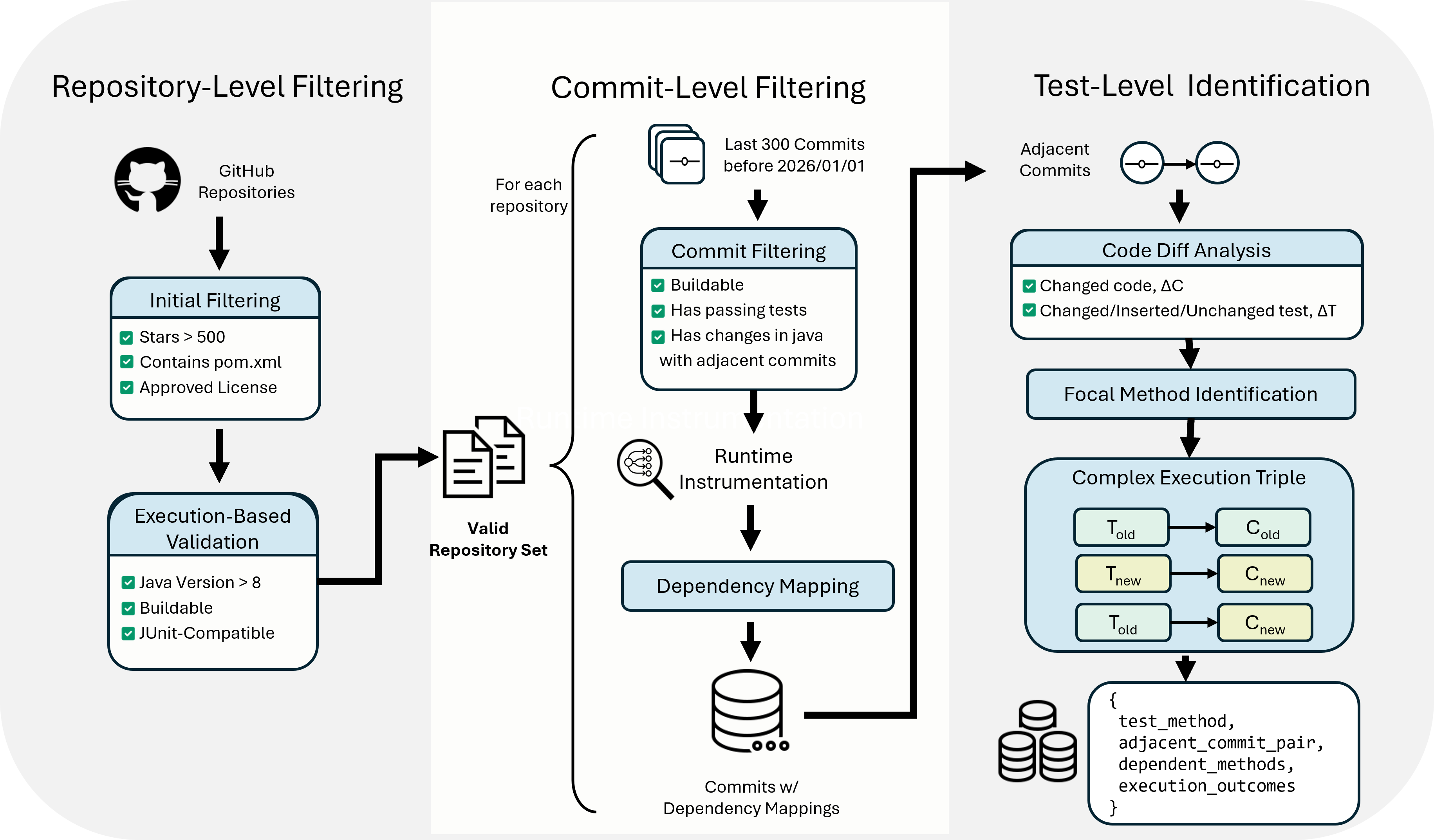}
\caption{Three-phase mining pipeline used to build \Bench: repository-level filtering, commit-level filtering with runtime dependency mapping, and test-level identification of code/test co-evolution triples.}
\label{fig:mining-details}
\end{figure}

Figure~\ref{fig:mining-details} illustrates the three-phase pipeline that turns a raw set of GitHub repositories into the per-task records consumed by the evaluation harness.

\MyPara{Repository-level filtering}
\Bench starts from the GitHub Search API, restricted to public Java repositories that contain a top-level \texttt{pom.xml} (so the Maven build harness applies), have at least $500$ stars, and use an OSI-approved license.
The search yields $2{,}262$ candidate repositories.
After cloning, each candidate is validated by execution: it must declare Java\,$\geq$\,8 in its Maven configuration, build cleanly with \texttt{mvn -DskipTests} from a recent commit on the default branch, and execute its existing JUnit suite without infrastructure errors.
Candidates that fail any of these checks are dropped, leaving $248$ repositories that feed the next phase.

\MyPara{Commit-level filtering}
For each retained repository, the pipeline walks the most recent $300$ commits before 2026-01-01 and forms adjacent commit pairs.
Each commit pair is kept only if both endpoints build cleanly, both have passing tests, and the pair contains Java code or test changes; commits that introduce only documentation, build-script, or configuration edits are filtered out at this stage.
For each surviving commit pair, \Bench instruments the project at test runtime to record per-test method-level dependencies (which production methods each test method actually exercises).
The instrumentation output is written as a dependency map alongside the diff metadata.
This yields buildable, test-passing, dependency-mapped commit pairs from $166$ repositories that feed the test-level identification step.

\MyPara{Test-level identification}
The final phase processes one (old revision, new revision, \texttt{test\_method}) triple at a time.
For every test method that exists in either revision, \Bench computes the method-level diff between the two revisions, classifies the test edit as \emph{changed}, \emph{inserted}, or \emph{unchanged}, and pairs it with the matching code diff $\Delta C$.
Each test is then linked to a focal production method using the dependency map collected in the previous phase, scored by static coupling so the strongest dependency wins.
\Bench then executes up to four cross-revision triples per task -- \OldOnOld, \NewOnNew, \OldOnNew, and \NewOnOld -- to record the behavioral fingerprint of each revision pair.
The triples drive task-type assignment: an \emph{update} task requires the old test to pass on the old code and the new test to pass on the new code, while the old test fails or fails to compile on the new code; a \emph{generation} task requires the new test to pass on the new code, and is further classified as \emph{discriminating} when the new test also fails on the old code.
The final per-repository JSONL records keep the test method, the adjacent commit pair, the dependent code methods, the focal method, and the recorded execution outcomes; \Bench packages these records into the per-track files consumed by the runner.
The released benchmark covers $91$ repositories that contribute at least one task to the test-generation or test-update track.

\subsection{Test Execution Environment}
\label{sec:appendix:details:environment}

\MyPara{Maven harness}
\Bench treats Maven as the unit of build, classpath, and test execution.
For each task, the harness checks out the new revision, applies the candidate test edit, and invokes \texttt{mvn -pl <module> -am test-compile} to recompile only the affected module and its build-time dependencies; this scopes recompilation to the test's module and avoids spurious failures in unrelated modules of multi-module projects.
Build plugins that are not relevant to compilation or test execution -- including \texttt{maven-checkstyle-plugin}, \texttt{spotbugs-maven-plugin}, \texttt{license-maven-plugin}, and similar style checkers -- are disabled at invocation time so that benign style violations introduced by an agent's edit do not mask a genuinely passing test.
The harness reads the compile-time classpath through \texttt{maven-dependency-plugin} (\texttt{dependency:build-classpath}) and reuses it across the compile and execute phases of the same task.

\MyPara{Test execution and isolation}
Each target test method is executed in isolation through the JUnit Platform launcher, with one fresh JVM per method, so that flakiness or static-state leaks across methods cannot improve or degrade a system's reported pass rate.
The harness records per-method outcomes (\emph{pass}, \emph{fail}, \emph{compile failure}, \emph{harness/code-injection failure}) along with stdout, stderr, and the JUnit XML report.
For each task, \Bench runs both the new test on the new code (\NewOnNew) and the old test on the new code (\OldOnNew); the latter is used to verify the update precondition that the pre-change test no longer passes after the code change.
For generation tasks, the harness additionally runs the new test on the old code (\NewOnOld) so that the discriminating signal -- test passes on new code and fails on old code -- can be recorded without retraining the agent on the old revision.

\MyPara{Coverage and dependency analysis}
Coverage is collected with JaCoCo by attaching \texttt{jacocoagent.jar} via \texttt{-javaagent} to the test JVM.
After execution, the harness invokes \texttt{jacococli.jar report} to materialize an XML report and intersects the covered line set with the line ranges of the focal dependency methods recorded at task-construction time, so the reported coverage is over the lines of the changed dependency methods rather than the full project.
A separate static analysis pass computes the focal-method dependency closure for each test, ranks dependencies by coupling, and stores the ranked list in the task spec; this list is the same one used during mining and is what coverage and mutation are evaluated against.
Coverage and mutation are reported only for methods whose tests passed: a failing test either short-circuits before exercising the focal code or instruments unrelated paths, and would yield misleading numbers either way.

\subsection{Agent Prompts}
\label{sec:appendix:prompts}

All four configurations share a single prompt string per track, assembled by a common prompt builder in our harness so that the four agents see byte-identical instructions.
\ClaudeCode and \GeminiCLI receive the string as the user message via their SDKs; \SWEAgent is configured to forward the same string verbatim as the task instance, with no system-level preamble or harness-specific wrapper.
Placeholders below in angle brackets are substituted at task-construction time from the task file: \verb|<commit_sha>| is the new revision, \verb|<test_file>| is the target test path, \verb|<test_methods>| lists the target methods, \verb|<deps_section>| is the deduplicated, stable-sorted \(\Delta M\) bullet list with each entry labeled as \emph{inserted}, \emph{modified}, or \emph{removed}, \verb|<cd_prefix>| is \texttt{cd <module> \&\&} for multi-module repos (empty otherwise), and \verb|<junit_jar>|, \verb|<junit_method_args>| configure the JUnit Console Launcher invocation.
We render \(\Delta M\) inside the listings as \texttt{DM} because the default \texttt{lstlisting} font has no Unicode coverage; the agent receives the verbatim string \texttt{DM} as well.

\MyPara{Test update prompt}

\begin{lstlisting}[basicstyle=\scriptsize\ttfamily,frame=single,framesep=2pt,xleftmargin=2pt,numbers=none,breaklines=true,columns=fullflexible,upquote=true]
You are updating JUnit tests to reflect a production code change in this repository.

REPOSITORY STATE:
- Pinned commit: <commit_sha>
- The production code in this commit reflects the POST-change state.
- The test file <test_file> currently reflects the PRE-change state and needs to be updated.

TARGET TESTS:
- File: <test_file>
- Methods to update:
<test_methods>

PRODUCTION METHODS CHANGED IN THIS COMMIT (DM):
<deps_section>

YOUR TASK:
1. Read the current test methods and the changed production methods.
2. Update each listed test method so it compiles against the new production code and passes.
3. The tests must exercise the same behavior as before where still relevant, and adapt to any new behavior introduced in DM.
4. Do NOT modify production code. Only modify <test_file>.
5. When done, ensure the tests compile and pass with the JUnit Console Launcher:
- Compile: `<cd_prefix>mvn test-compile`
- Run with the JUnit Console Launcher (jar at `<junit_jar>`). Set `--class-path` to include the project's compiled classes and dependencies, then invoke:
`<cd_prefix>java -jar <junit_jar> --class-path "<classpath>" --reports-dir ./test-reports <junit_method_args>`
6. If `MAVEN_SETTINGS_PATH` is set in the shell, pass `-s "$MAVEN_SETTINGS_PATH"` to Maven commands so dependency downloads use the configured cache.
7. Before finishing, run `git diff --name-only HEAD`. If any tracked file other than <test_file> changed, especially due to Maven/license-header side effects, revert those non-target files with `git checkout -- <path>`. The final diff must contain only <test_file>.

You have file read/edit and shell tools. Work autonomously; do not ask for confirmation.
\end{lstlisting}

\MyPara{Test generation prompt}

\begin{lstlisting}[basicstyle=\scriptsize\ttfamily,frame=single,framesep=2pt,xleftmargin=2pt,numbers=none,breaklines=true,columns=fullflexible,upquote=true]
You are writing new JUnit test methods for a production code change in this repository.

REPOSITORY STATE:
- Pinned commit: <commit_sha>
- The production code in this commit reflects the POST-change state.
- The test file <test_file> does NOT yet contain tests for this change; you must add them.

TARGET TESTS:
- File: <test_file>
- Methods to add:
<test_methods>

PRODUCTION METHODS CHANGED IN THIS COMMIT (DM):
<deps_section>

YOUR TASK:
1. Read the changed production methods to understand the new behavior.
2. For each method listed under "Methods to add", write a JUnit test method with that exact name in <test_file> that exercises the new behavior introduced in DM. Annotate with @Test and return void.
3. Each generated test must be discriminating: it should PASS on this POST-change code, but would FAIL or fail-to-compile if the production code were reverted to the PRE-change version. Avoid tests that only assert unchanged behavior.
4. If <test_file> does not exist, create it with the appropriate package declaration and imports.
5. Do NOT modify production code. Only modify <test_file>.
6. When done, ensure the tests compile and pass with the JUnit Console Launcher:
- Compile: `<cd_prefix>mvn test-compile`
- Run with the JUnit Console Launcher (jar at `<junit_jar>`). Set `--class-path` to include the project's compiled classes and dependencies, then invoke:
`<cd_prefix>java -jar <junit_jar> --class-path "<classpath>" --reports-dir ./test-reports <junit_method_args>`
7. If `MAVEN_SETTINGS_PATH` is set in the shell, pass `-s "$MAVEN_SETTINGS_PATH"` to Maven commands so dependency downloads use the configured cache.
8. Before finishing, run `git diff --name-only HEAD`. If any tracked file other than <test_file> changed, especially due to Maven/license-header side effects, revert those non-target files with `git checkout -- <path>`. The final diff must contain only <test_file>.

You have file read/edit and shell tools. Work autonomously; do not ask for confirmation.
\end{lstlisting}

\noindent
An example \texttt{<deps\_section>} for the update task in Figure~\ref{fig:task-examples:update}:
\begin{lstlisting}[basicstyle=\scriptsize\ttfamily,frame=single,framesep=2pt,xleftmargin=2pt,numbers=none,breaklines=true,columns=fullflexible,upquote=true]
- **modified** `public String toString()' in `bxbot-core/src/main/java/com/gazbert/bxbot/core/engine/BotStatus.java`
\end{lstlisting}

\subsection{Coverage Analysis}
\label{sec:appendix:details:coverage}

\MyPara{Pipeline}
Coverage is collected with JaCoCo~\citep{JaCoCo}.
\Bench attaches \texttt{jacocoagent.jar} via \texttt{-javaagent} during \texttt{mvn test}, producing \texttt{jacoco.exec}; \texttt{jacococli.jar} then renders an XML report at \texttt{coverage.xml}.
Coverage is computed only when the test under evaluation passes -- a failing test either short-circuits before exercising the focal code or instruments unrelated paths, so the resulting number would be misleading.

\MyPara{Focal line coverage}
The reported coverage number is the line coverage of the test under evaluation on the focal dependency methods of the task.
Each dep method carries a \texttt{file\_path} and a line range \texttt{line\_nums} (e.g.\ \texttt{125-131}).
For every line $\ell$ in that range, \Bench locates the matching \texttt{<sourcefile>} entry in the JaCoCo XML and counts $\ell$ as covered iff its \texttt{ci} attribute (covered instructions on $\ell$) is positive.
Aggregated across all dep methods of the task,
\begin{equation}
\mathtt{line\_coverage} \;=\; \frac{\sum_{\text{deps}} \mathtt{covered\_lines}}{\sum_{\text{deps}} \mathtt{total\_lines}} \times 100.
\label{eq:focal-line-cov}
\end{equation}
This same formula is applied twice per method-eval entry in the result JSON, producing two flat fields:
\texttt{coverage\_pred} -- line coverage of the system-generated test, recomputed live by re-running the test with JaCoCo on the new code; and
\texttt{coverage\_gold} -- line coverage of the developer's gold test, precomputed at data-construction time and read from the task spec.
The asymmetry is intentional: the gold test is fixed across all submissions, so re-running it for every system would be redundant work; the predicted test changes per system and must therefore be recomputed live.
Branch and method coverage are not used in the leaderboard.

\subsection{Mutation Analysis}
\label{sec:appendix:details:mutation}

\MyPara{Mutant generation}
For each focal dependency method, \Bench generates mutants on the new code revision using Universal Mutator~\citep{GroceETAL18UniversalMutator}.
UM is invoked per Java source file with a \texttt{--lines} filter restricted to the intersection of the dep method's body and the lines added in the diff between the old and new revisions, so mutants only target code that the new revision introduced or modified.
A per-file cap of $10$ mutants is applied through seeded random sampling, which keeps scoring runs deterministic and bounds the per-task runtime when UM proposes large numbers of mutants on long methods.
The cap is shared across all dep methods that live in the same Java file: if three dep methods of one task all reside in \texttt{Foo.java}, their pooled candidate mutants are sampled down to ten total, not ten per dep method.

\MyPara{Mutant accounting}
Mutants are generated per test method: for a given test method, \Bench mutates every dependency method on its dep list and then runs that single test method against the pooled mutant set to obtain its mutation score, so all per-method counts in Table~\ref{tab:mutation-method-stats} (\#Methods w/ Mutants, \#Generated/Method, \#Killed/Method) are taken over test methods rather than over dep methods or files.
We partition UM's output into five accounting categories so that no mutant is silently dropped:
\begin{itemize}
\item \texttt{emitted} -- total mutants UM produced for the file (UM's VALID, INVALID, and REDUNDANT outputs combined). This equals the sum of the four categories below.
\item \texttt{compile\_failed} -- UM's INVALID: a mutant whose source no longer compiles under \texttt{javac}, and so cannot serve as a kill target.
\item \texttt{redundant} -- UM's REDUNDANT: a mutant identical to the original source or to another emitted mutant.
\item \texttt{capped\_out} -- compile-valid, non-redundant mutants discarded by the per-file cap of $10$. Non-zero only when UM produced more than $10$ such mutants for the file. This is the only post-filtering \Bench performs on UM's output.
\item \texttt{count} -- final mutants written to disk and used as the kill-set for scoring.
\end{itemize}
These counts satisfy the invariant
\begin{equation}
\mathtt{emitted} \;=\; \mathtt{count} + \mathtt{compile\_failed} + \mathtt{redundant} + \mathtt{capped\_out},
\label{eq:mutant-invariant}
\end{equation}
which lets a reader distinguish ``UM produced few mutants'' from ``the per-file cap is biting'' from ``many mutants do not compile under the project's classpath''.

\MyPara{Notation}
Let $M_n$ be the kept mutant set on the new code (its cardinality $|M_n|$ is the \texttt{count} value above).
For a test $T$ run against $M_n$, $\mathtt{killed}_T$ is the number of mutants killed -- equivalently, the number of mutants for which $T$ does not return \texttt{pass}. Both $T = \text{gold}$ (the developer's reference test on the new revision) and $T = \text{pred}$ (the system-generated test on the new revision) are scored against the same $M_n$, so the two values are directly comparable.

\MyPara{Mutation scores}
For each task, $\mathtt{killed}_{\text{gold}}$ (the gold new test's kill count on new-code mutants) and $|M_n|$ are taken from the mutation results precomputed at task-construction time and released with the dataset; the predicted test is then run against the same cached new-code mutants to obtain $\mathtt{killed}_{\text{pred}}$.
The per-method scores are then
\begin{align}
\mathtt{mutation\_score\_gold} &= \mathtt{killed}_{\text{gold}} \,/\, |M_n|, \label{eq:headline-gold} \\
\mathtt{mutation\_score\_pred} &= \mathtt{killed}_{\text{pred}} \,/\, |M_n|. \label{eq:headline-pred}
\end{align}
Both scores divide by the same $|M_n|$, so they lie in $[0,1]$ and share a denominator within a method, which is why a per-method comparison of pred against gold is meaningful.
A per-method score is set to $\bot$ when $|M_n| = 0$, when the cached gold counts are missing, or (for \texttt{mutation\_score\_pred}) when the predicted test failed to compile or run.
Per-method $\bot$ values are preserved at the method level rather than coerced to zero so that the per-task aggregation step (Sec.~\ref{sec:appendix:details:aggregation}) can distinguish ``ran and killed nothing'' from ``did not produce a measurement.''

\begin{table}[t]
\caption{Per-test-method mutation statistics for each track. \emph{\#Test Methods\,[\#Tasks]} reports the exact benchmark count of test methods, with the number of underlying tasks in brackets. \emph{\#Methods w/ Mutants}, \emph{\#Generated/Method}, and \emph{\#Killed/Method} are averaged across the four agent result files per track: \emph{\#Methods w/ Mutants} is the mean number of test methods for which a mutation tool generated at least one mutant, with the percentage taken over \#Test Methods; \emph{\#Generated/Method} and \emph{\#Killed/Method} report the mean per-method counts of generated and killed mutants over those methods, and the percentage on \emph{\#Killed/Method} is the kill rate \#Killed/\#Generated.}
\label{tab:mutation-method-stats}
\centering
\small
\resizebox{\linewidth}{!}{%

\begin{tabular}{lrrrr}
\toprule
\textbf{Track} & \textbf{\#Test Methods\,[\#Tasks]} & \textbf{\#Methods w/ Mutants} & \textbf{\#Generated/Method} & \textbf{\#Killed/Method} \\
\midrule
Test generation & \UseMacro{MutGen-TotalMethods}\,[\UseMacro{MutGen-TotalTasks}] & \UseMacro{MutGen-MethodsWithMutants} (\UseMacro{MutGen-MethodsWithMutantsPct}) & \UseMacro{MutGen-GeneratedPerMethod} & \UseMacro{MutGen-KilledPerMethod} (\UseMacro{MutGen-KilledPerMethodPct}) \\
Test update & \UseMacro{MutUpdate-TotalMethods}\,[\UseMacro{MutUpdate-TotalTasks}] & \UseMacro{MutUpdate-MethodsWithMutants} (\UseMacro{MutUpdate-MethodsWithMutantsPct}) & \UseMacro{MutUpdate-GeneratedPerMethod} & \UseMacro{MutUpdate-KilledPerMethod} (\UseMacro{MutUpdate-KilledPerMethodPct}) \\
\bottomrule
\end{tabular}

}
\end{table}

\subsection{Aggregation Across Methods and Tasks}
\label{sec:appendix:details:aggregation}

Both mutation score and focal line coverage are first computed at the level of a single test method, then aggregated to the task level, and finally averaged across tasks.
The result JSON exposes two flavors of the per-task average for each metric.
We give the construction here once because mutation and coverage share it.

\MyPara{Per-method score}
For mutation, a method's score is either \texttt{mutation\_score\_gold} or \texttt{mutation\_score\_pred} from Eqs.~\ref{eq:headline-gold}--\ref{eq:headline-pred} (set to $\bot$ when $|M_n| = 0$, when the cached gold counts are missing, or, for the pred variant, when the predicted test did not pass).
For coverage, a method's score is the focal line coverage from Eq.~\ref{eq:focal-line-cov} for the predicted test, or the precomputed value for the gold test ($\bot$ if the predicted test did not pass or the gold value is missing).

\MyPara{Per-task average: wide vs.\ narrow denominator}
Let $N$ be the total number of test methods on the task, and let $S$ be the count of those methods for which the metric was successfully computed (the per-method score is not $\bot$).
Per-method $\bot$ values are preserved at the method level so consumers can distinguish ``ran and scored $0$'' from ``did not score at all'', but at the per-task aggregation step the two reported flavors handle them differently.
Define the wide-form numerator $X_{\text{wide}}$ as the sum of per-method scores with $\bot$ folded to $0$, and the narrow-form numerator $X_{\text{narrow}}$ as the sum over only the $S$ methods that produced a measurement; the result JSON reports
\begin{equation}
\mathtt{avg\_X} \;=\; \frac{X_{\text{wide}}}{N},
\qquad
\mathtt{avg\_X\_success} \;=\; \frac{X_{\text{narrow}}}{S}.
\label{eq:wide-narrow}
\end{equation}
The wide form treats a method without a measurement as a score of $0$ -- conflating ``the test ran but killed no mutants / covered no lines'' with ``no measurement was produced'' -- but penalises systems that fail to produce measurements.
The narrow form restricts both numerator and denominator to methods that actually produced a measurement, isolating per-method quality from coverage-over-attempts.
The two flavors are not weighted means; they are arithmetic means with different denominators, reported jointly so a reader can separate per-method quality (narrow) from coverage-over-attempts (wide).

\MyPara{Cross-task aggregate}
The headline numbers in Section~\ref{sec:experiments} are the arithmetic mean of the per-task averages across tasks for which the per-task average is defined.
Each task contributes a single value, so tasks are weighted equally regardless of how many test methods they contain; this is a deliberate choice to prevent a small number of large tasks from dominating the leaderboard, at the cost of giving a one-method task the same weight as a fifty-method task.
The same construction is applied to all four quantities reported by the runner -- $\mathtt{coverage\_pred}$, $\mathtt{coverage\_gold}$, $\mathtt{mutation\_score\_gold}$, and $\mathtt{mutation\_score\_pred}$ -- each in both wide and \texttt{\_success} form, for eight leaderboard fields total.

\subsection{Codebase Statistics}
\label{sec:appendix:details:codebase}

\begin{table}[t]
\caption{\Bench codebase summary. Per-repository means are computed across the $91$ repositories that contribute at least one task to the test-generation or test-update track. Number of test methods is the per-task mean averaged over repositories. Java-file and Java-line means are taken over the repositories with a successful repository scan.}
\label{tab:bench-codebase-overview}
\centering
\small

\begin{tabular}{llr}
\toprule
& \textbf{Metric} & \textbf{Value} \\
\midrule
\multirow{6}{*}{Codebase}
& Repositories          & 91 \\
& \# Commits (mean)      & 7 \\
& \# Tasks (mean)        & 14 \\
& \# Test methods (mean) & 2 \\
& \# Java files (mean)   & 214 \\
& \# Java lines (mean)   & 17{,}049 \\
\bottomrule
\end{tabular}

\end{table}

Table~\ref{tab:bench-codebase-overview} summarises the size of the $91$ repositories that contribute at least one task to the current \Bench snapshot.
On average each repository spans $214$ Java files and $17{,}049$ lines of Java, and supplies $7$ commit pairs and $14$ tasks, with $2$ test methods per task on average.
These are substantial codebases rather than isolated snippets, so a successful agent must navigate non-trivial source, dependencies, and build configuration in addition to producing the correct test edit; this is also reflected in the per-task token budgets reported in Table~\ref{tab:benchmark-track-statistics}.

\subsection{Hardware Environment}
\label{sec:appendix:details:hardware}

The benchmark construction and experiments were conducted on a cluster of virtual machines, each with 4 CPUs, 15GB RAM, and 103GB disk space.
Each virtual machine runs the tests from one repository/task at a time, while an orchestrator manages the parallel execution of the repositories/tasks.

\section{Additional Experimental Results}
\label{sec:appendix:additional-results}

This appendix expands the on-pass quality columns of Tables~\ref{tab:test-generation-results} and~\ref{tab:test-update-results} into the full coverage and mutation breakdown.
Tables~\ref{tab:cov-detail-generation}--\ref{tab:mut-detail-update} report, for every (track, agent) pair, four numbers per metric: the wide-form value (denominator $=$ all configured methods, with non-measured methods folded to $0$); the on-pass value (denominator $=$ methods whose tests passed); the developer-test reference on those same passing methods; and the signed pred-minus-gold delta on the on-pass set, so a reader can see how far each agent-produced test is from the developer reference on the methods it managed to pass.
Table~\ref{tab:cost-cutoff-results} then reports the cost-limited operating point analyzed in Section~\ref{sec:experiments:cost}.

\begin{table}[t]
\caption{Coverage detail for the test-generation track. \CovMetric\ is the wide-form line coverage on the focal dependency methods (denominator $=$ all methods); \CovOnPassMetric\ averages only over methods whose tests passed; \CovOnPassHumanMetric\ is the developer-written test's coverage on those same methods; \DeltaCovOnPassMetric\ is the signed pred-minus-gold difference.}
\label{tab:cov-detail-generation}
\centering
\small
\resizebox{\linewidth}{!}{%

\begin{tabular}{llrrrr}
\toprule
\textbf{\UseMacro{TH-Harness}} & \textbf{\UseMacro{TH-Model}} & \textbf{\UseMacro{TH-Cov}\,$\uparrow$} & \textbf{\UseMacro{TH-CovOnPass}\,$\uparrow$} & \textbf{\UseMacro{TH-CovOnPassHuman}} & \textbf{\UseMacro{TH-DeltaCovOnPass}} \\
\midrule
\ClaudeCode & \ClaudeOpus & \textbf{74.8\%} & 76.8\% & 80.8\% & -4.043\% \\
\GeminiCLI & \GeminiPro & 73.2\% & 75.1\% & 80.8\% & -5.699\% \\
\SWEAgent & \ClaudeOpus & 65.1\% & \textbf{78.0\%} & 80.5\% & -2.518\% \\
\SWEAgent & \GeminiPro & 65.3\% & 74.3\% & 80.4\% & -6.100\% \\
\bottomrule
\end{tabular}

}
\end{table}

\begin{table}[t]
\caption{Mutation detail for the test-generation track. \MutMetric\ is the wide-form mutation kill ratio (denominator $=$ all methods); \MutOnPassMetric\ averages only over methods whose tests passed; \MutOnPassHumanMetric\ is the developer-written test's kill ratio on those same methods; \DeltaMutOnPassMetric\ is the signed pred-minus-gold difference.}
\label{tab:mut-detail-generation}
\centering
\small
\resizebox{\linewidth}{!}{%

\begin{tabular}{llrrrr}
\toprule
\textbf{\UseMacro{TH-Harness}} & \textbf{\UseMacro{TH-Model}} & \textbf{\UseMacro{TH-Mut}\,$\uparrow$} & \textbf{\UseMacro{TH-MutOnPass}\,$\uparrow$} & \textbf{\UseMacro{TH-MutOnPassHuman}} & \textbf{\UseMacro{TH-DeltaMutOnPass}} \\
\midrule
\ClaudeCode & \ClaudeOpus & \textbf{40.6\%} & 56.6\% & 54.2\% & +2.338\% \\
\GeminiCLI & \GeminiPro & 39.3\% & 55.0\% & 54.2\% & +0.800\% \\
\SWEAgent & \ClaudeOpus & 34.9\% & \textbf{57.1\%} & 55.3\% & +1.791\% \\
\SWEAgent & \GeminiPro & 35.8\% & 55.6\% & 54.4\% & +1.110\% \\
\bottomrule
\end{tabular}

}
\end{table}

\begin{table}[t]
\caption{Coverage detail for the test-update track. Columns match Table~\ref{tab:cov-detail-generation}.}
\label{tab:cov-detail-update}
\centering
\small
\resizebox{\linewidth}{!}{%

\begin{tabular}{llrrrr}
\toprule
\textbf{\UseMacro{TH-Harness}} & \textbf{\UseMacro{TH-Model}} & \textbf{\UseMacro{TH-Cov}\,$\uparrow$} & \textbf{\UseMacro{TH-CovOnPass}\,$\uparrow$} & \textbf{\UseMacro{TH-CovOnPassHuman}} & \textbf{\UseMacro{TH-DeltaCovOnPass}} \\
\midrule
\ClaudeCode & \ClaudeOpus & \textbf{59.1\%} & \textbf{79.4\%} & 80.0\% & -0.585\% \\
\GeminiCLI & \GeminiPro & \textbf{59.1\%} & 79.1\% & 80.0\% & -0.876\% \\
\SWEAgent & \ClaudeOpus & 52.0\% & 79.2\% & 80.5\% & -1.257\% \\
\SWEAgent & \GeminiPro & 58.4\% & 79.1\% & 79.9\% & -0.846\% \\
\bottomrule
\end{tabular}

}
\end{table}

\begin{table}[t]
\caption{Mutation detail for the test-update track. Columns match Table~\ref{tab:mut-detail-generation}.}
\label{tab:mut-detail-update}
\centering
\small
\resizebox{\linewidth}{!}{%

\begin{tabular}{llrrrr}
\toprule
\textbf{\UseMacro{TH-Harness}} & \textbf{\UseMacro{TH-Model}} & \textbf{\UseMacro{TH-Mut}\,$\uparrow$} & \textbf{\UseMacro{TH-MutOnPass}\,$\uparrow$} & \textbf{\UseMacro{TH-MutOnPassHuman}} & \textbf{\UseMacro{TH-DeltaMutOnPass}} \\
\midrule
\ClaudeCode & \ClaudeOpus & 19.2\% & 44.6\% & 46.2\% & -1.614\% \\
\GeminiCLI & \GeminiPro & \textbf{19.4\%} & 44.9\% & 46.4\% & -1.448\% \\
\SWEAgent & \ClaudeOpus & 16.8\% & \textbf{46.0\%} & 48.0\% & -2.015\% \\
\SWEAgent & \GeminiPro & 19.1\% & 44.7\% & 46.3\% & -1.620\% \\
\bottomrule
\end{tabular}

}
\end{table}

\clearpage

\begin{table*}[!htbp]
\caption{Cost-limited operating point on the test-generation and test-update tracks. Each (track, harness, model) is shown at the default per-task budget (\$3) and at two tighter caps (\$1, \$0.5). Tasks whose billed cost exceeds the cap are treated as code-injection failures and counted in \HrnsFailPct; at the default-budget row, \HrnsFailPct\ is the native harness failure rate from \texttt{agent\_run\_info.exit\_status}. \UseMacro{TH-SuccessPct} is the fraction of target methods whose tests pass on the new revision and (when recorded) fail on the old revision, with the full method count as denominator. Avg Tokens / Avg Cost are means over tasks at or below the budget.}
\label{tab:cost-cutoff-results}
\centering
\small
\resizebox{\linewidth}{!}{%

\begin{tabular}{llllrrrr}
\toprule
\textbf{Track} & \textbf{\UseMacro{TH-Harness}} & \textbf{\UseMacro{TH-Model}} & \textbf{Budget} & \textbf{Avg Tokens} & \textbf{Avg Cost} & \textbf{\UseMacro{TH-SuccessPct}} & \textbf{HrnsFail\%} \\
\midrule
\multirow{12}{*}{test-generation} & \multirow{3}{*}{\ClaudeCode} & \multirow{3}{*}{\ClaudeOpus} & \$3 & 8.6k & \$0.88 & 70.6\% & 0.0\% \\
&  &  & \$1 & 6.0k & \$0.57 & 44.2\% & 30.7\% \\
&  &  & \$0.5 & 4.2k & \$0.37 & 21.6\% & 69.8\% \\
\cmidrule(lr){2-8}
& \multirow{3}{*}{\GeminiCLI} & \multirow{3}{*}{\GeminiPro} & \$3 & 90.0k & \$0.33 & 71.3\% & 0.0\% \\
&  &  & \$1 & 85.7k & \$0.31 & 69.8\% & 2.1\% \\
&  &  & \$0.5 & 73.6k & \$0.25 & 59.3\% & 16.2\% \\
\cmidrule(lr){2-8}
& \multirow{6}{*}{\SWEAgent} & \multirow{3}{*}{\ClaudeOpus} & \$3 & 182.9k & \$1.77 & 59.5\% & 22.8\% \\
&  &  & \$1 & 59.4k & \$0.70 & 18.8\% & 70.8\% \\
&  &  & \$0.5 & 32.5k & \$0.44 & 3.0\% & 96.2\% \\
\cmidrule(lr){3-8}
&  & \multirow{3}{*}{\GeminiPro} & \$3 & 896.3k & \$0.49 & 63.5\% & 2.5\% \\
&  &  & \$1 & 591.1k & \$0.35 & 59.1\% & 9.0\% \\
&  &  & \$0.5 & 394.1k & \$0.28 & 49.4\% & 26.0\% \\
\midrule
\multirow{12}{*}{test-update} & \multirow{3}{*}{\ClaudeCode} & \multirow{3}{*}{\ClaudeOpus} & \$3 & 9.2k & \$1.18 & 86.1\% & 0.0\% \\
&  &  & \$1 & 5.8k & \$0.59 & 54.8\% & 44.6\% \\
&  &  & \$0.5 & 4.0k & \$0.37 & 14.4\% & 78.2\% \\
\cmidrule(lr){2-8}
& \multirow{3}{*}{\GeminiCLI} & \multirow{3}{*}{\GeminiPro} & \$3 & 157.8k & \$0.58 & 86.6\% & 0.0\% \\
&  &  & \$1 & 106.8k & \$0.39 & 85.8\% & 13.8\% \\
&  &  & \$0.5 & 85.1k & \$0.30 & 62.9\% & 33.8\% \\
\cmidrule(lr){2-8}
& \multirow{6}{*}{\SWEAgent} & \multirow{3}{*}{\ClaudeOpus} & \$3 & 211.4k & \$2.03 & 73.2\% & 40.3\% \\
&  &  & \$1 & 62.9k & \$0.73 & 18.1\% & 74.9\% \\
&  &  & \$0.5 & 34.5k & \$0.46 & 1.1\% & 97.8\% \\
\cmidrule(lr){3-8}
&  & \multirow{3}{*}{\GeminiPro} & \$3 & 1551.0k & \$0.82 & 86.0\% & 12.0\% \\
&  &  & \$1 & 719.1k & \$0.41 & 79.9\% & 19.4\% \\
&  &  & \$0.5 & 485.7k & \$0.32 & 58.3\% & 40.1\% \\
\bottomrule
\end{tabular}

}
\end{table*}

Across Tables~\ref{tab:cov-detail-generation} and~\ref{tab:cov-detail-update}, the wide-to-on-pass gap on coverage is dominated by the outcome-category distribution: configurations with lower \SuccessPct\ have correspondingly lower \CovMetric\ even though their \CovOnPassMetric\ is on par with the higher-\SuccessPct\ configurations.
The reference \CovOnPassHumanMetric\ is around $80\%$ on both tracks, which suggests that focal-line coverage is bounded above by the dependency-method line ranges that the developer test was already targeting; agent-produced tests reach about three quarters of that ceiling on the update track and within a few percentage points of it on the generation track.
For mutation, Tables~\ref{tab:mut-detail-generation} and~\ref{tab:mut-detail-update} show that \MutOnPassMetric\ matches or slightly exceeds \MutOnPassHumanMetric\ on every test-generation row, indicating that on the methods whose tests passed, the agent-produced tests kill at least as many of the cached new-code mutants as the developer-written tests do.

\section{Replication Study: Test Update Identification with Dynamic Labels}
\label{sec:appendix:test-update-identification}

\begin{figure}[t]
\centering
\includegraphics[width=.55\linewidth]{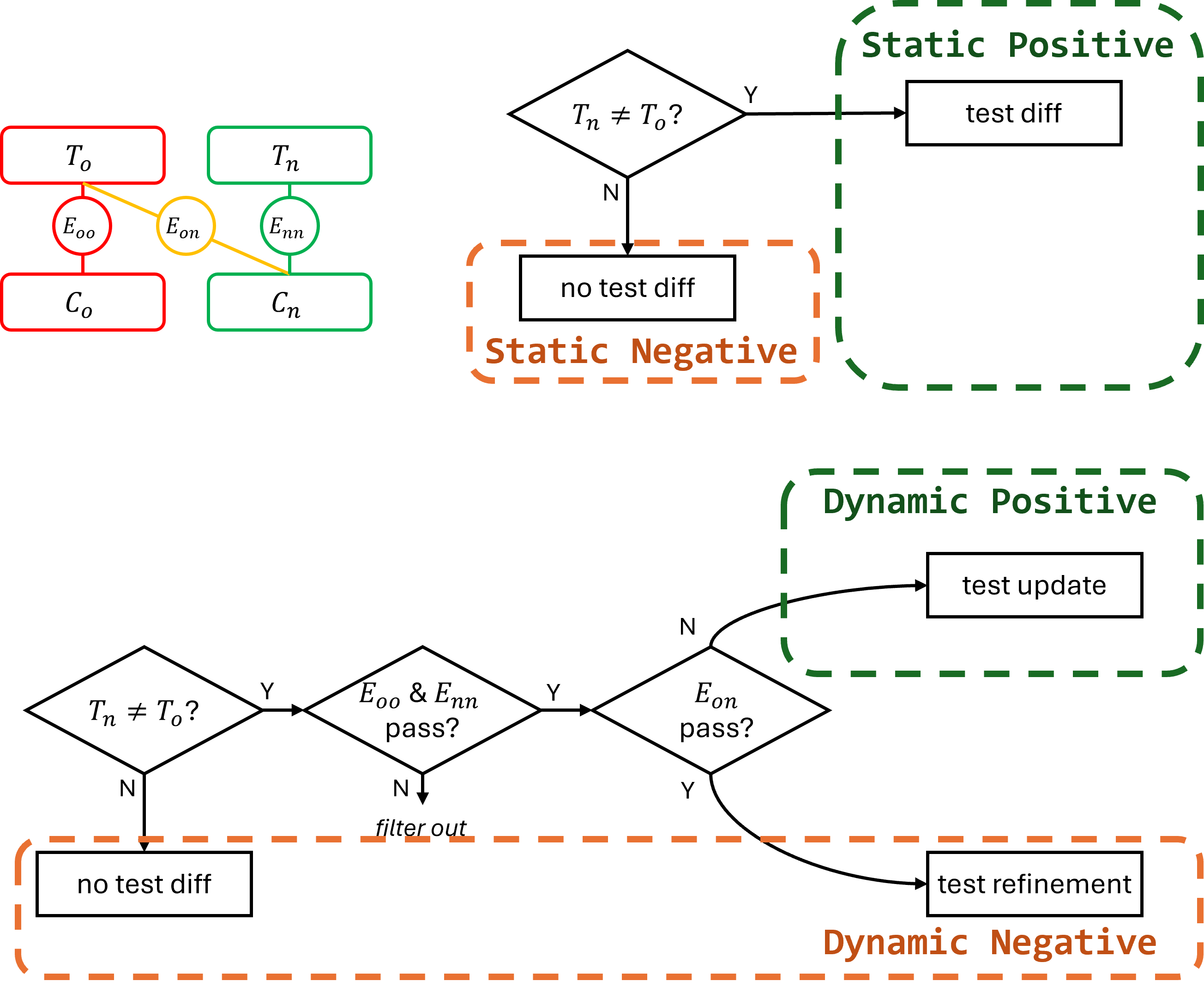}
\caption{Static versus execution-backed (``dynamic'') labelling of test-update positives. Top: a static heuristic flags any commit in which the test changes ($T_n \neq T_o$) as a positive. Bottom: the dynamic label additionally requires the old test to pass on the old code ($E_{oo}$) and the new test to pass on the new code ($E_{nn}$), then promotes a candidate to a positive only when the old test fails or fails to compile on the new code ($E_{on}$); commits whose tests change but pass under all three executions are classified as refinements and excluded.}
\label{fig:test-update-identification}
\end{figure}

The main paper de-prioritises test-update identification as a benchmark track, but the analysis remains important evidence for why \Bench uses execution-backed labels.
Prior work such as CEPROT~\cite{HuETAL23CEPROT} labels a test as obsolete when code and test co-change in the same commit.
We compare that static heuristic with the execution-backed criterion summarised in Figure~\ref{fig:test-update-identification}: the old test must pass on the old code, the new test must pass on the new code, and the old test must fail or fail to compile on the new code.
The replication study runs on a smaller subset of $105$ repositories chosen to overlap with the CEPROT evaluation set; this is a strict subset of the \NumCandidateProjects\ post-mining repositories used by the main benchmark and is referenced only here.
Table~\ref{tab:rq1-replication-groups} reports the repository, task, and test-method counts for each label class under the two labelling regimes.

\begin{table}[t]
\caption{Replication-study label classes on the $105$-repository CEPROT-overlap subset. Static labels follow CEPROT's co-change heuristic; dynamic labels additionally require execution-confirmed obsolescence.}
\label{tab:rq1-replication-groups}
\centering
\small

\begin{tabular}{llrrr}
\toprule
\textbf{Labelling} & \textbf{Class} & \textbf{\#Repo} & \textbf{\#Task} & \textbf{\#Test Method} \\
\midrule
Static  & Positive &  66 &   627 &  1{,}369 \\
Static  & Negative &  99 & 2{,}603 & 10{,}496 \\
\midrule
Dynamic & Positive &  59 &   509 &  1{,}138 \\
Dynamic & Negative & 100 & 2{,}721 & 10{,}727 \\
\bottomrule
\end{tabular}

\end{table}

On this $105$-repository subset, the static heuristic identifies $1{,}369$ positive test methods while execution-backed labels confirm only $1{,}138$.
The $231$ static-only positives correspond to co-changed tests that are not confirmed as broken by execution; these are predominantly refinements, assertion strengthening, and optional coverage improvements rather than required updates.

\begin{table}[t]
\caption{Cross-dataset CEPROT evaluation results. Each row reports F1, precision, and recall for a model trained on one label source and evaluated on another.}
\label{tab:rq1-ceprot-crosseval}
\centering
\small

\begin{tabular}{llccc}
\toprule
\textbf{Train} & \textbf{Test} & \textbf{F1} & \textbf{Precision} & \textbf{Recall} \\
\midrule
Dynamic & Dynamic & 0.74 & 0.86 & 0.66 \\
Dynamic & Static  & 0.67 & 0.86 & 0.55 \\
\midrule
Static  & Dynamic & \textbf{0.79} & 0.85 & \textbf{0.74} \\
Static  & Static  & 0.68 & 0.79 & 0.60 \\
\bottomrule
\end{tabular}

\end{table}

Table~\ref{tab:rq1-ceprot-crosseval} reports a CEPROT cross-evaluation over the static and execution-backed label sources.
F1 ranges from $0.67$ to $0.79$ across train/test combinations.
The best result occurs when training on the broader static labels and evaluating on execution-backed ground truth; static-only evaluation, by contrast, remains problematic because it treats non-broken co-changes as required repairs.
This supports the benchmark design choice: static analysis is useful for candidate generation, while execution is necessary for accurate labels.

\section{Broader Impact}
\label{sec:appendix:broader-impact}

\MyPara{Positive impacts}
\Bench provides an executable, contamination-aware benchmark for LLM-driven test generation and test update, two tasks that occupy a substantial share of developer effort in industrial code maintenance.
By replacing diff-similarity scoring with build, test, coverage, and mutation execution, the benchmark gives practitioners and researchers a faithful signal for whether a generated or updated test compiles, exercises the intended code, and detects realistic faults, which supports better-informed decisions about deploying agentic coding tools in continuous-integration pipelines.
The accompanying open leaderboard and live ingestion of new revisions further narrow the gap between published numbers and the conditions under which these tools are used in practice.

\MyPara{Potential negative impacts and mitigations}
A tight executable benchmark can encourage configurations that overfit to its specific cleaning filters, producing agents that score well on \Bench but degrade on broader workloads.
We mitigate this by publishing the full mining and execution pipeline, reporting complementary metrics beyond pass rate (\RedundantPct, \CovOnPassMetric, \MutOnPassMetric), and continuously ingesting new tasks anchored to recent revisions so that any single \Bench snapshot is only one slice of an evolving evaluation.
A second concern is that a high \SuccessPct\ on \Bench may overstate readiness for production, because a generated test that compiles and passes can still encode a shallow oracle that hides regressions.
\Bench therefore reports coverage and mutation alongside pass rate and exposes per-task signals so adopters can audit individual agent outputs rather than relying on aggregate scores.

\end{document}